\documentclass[11pt]{iopart}
\expandafter\let\csname equation*\endcsname\relax
\expandafter\let\csname endequation*\endcsname\relax
\usepackage{amssymb}
\usepackage{amsmath}
\usepackage{bm}
\usepackage{color}
\usepackage{epsfig}
\usepackage{float}
\usepackage[numbers,sort&compress]{natbib}
\usepackage{hyperref} 

\begin{document}
\title[Minimal quantum thermal machine in a bandgap environment]{Minimal quantum thermal machine in a bandgap environment: non-Markovian features and anti-Zeno advantage}
\author{Meng Xu$^1$, J. T. Stockburger$^1$, G. Kurizki$^2$ and J. Ankerhold$^1$}
\address{$^1$ Institute  for Complex Quantum Systems and IQST, Ulm University - Albert-Einstein-Allee 11, D-89069  Ulm, Germany}
\address{$^2$ Department of Chemical and Biological Physics, Weizmann Institute of Science, Rehovot 7610001, Israel}
\ead{\mailto{meng.xu@uni-ulm.de}}
\vspace{10pt}
\begin{indented}
\item December 2021
\end{indented}

\begin{abstract}
A minimal model of a quantum thermal machine is analyzed, where a driven two level working medium  (WM) is embedded in an environment (reservoir) whose spectrum possesses bandgaps. The transition frequency of the WM is periodically modulated so as to be in alternating spectral overlap with hot or cold reservoirs whose spectra are separated by a bandgap. Approximate and exact treatments supported by analytical considerations yield a complete characterization of this thermal machine in the deep quantum domain. For slow to moderate modulation, the spectral response of the reservoirs is close to equilibrium, exhibiting sideband (Floquet) resonances in the heat currents and power output. In contrast, for faster modulation, strong-coupling and non-Markovian features give rise to correlations  between the WM and the reservoirs and between the two reservoirs. Power boost of strictly quantum origin (“quantum advantage”) is then found for both continuous and segmental fast modulation that leads to the anti-Zeno effect of enhanced spectral reservoir response. Such features cannot be captured by standard Markovian treatments.
\end{abstract}

\noindent{\it Keywords\/}: Quantum thermal machines, quantum heat transport, non-perturbative open quantum dynamics, anti-Zeno effect, spectral bandgaps

\section{Introduction}
The description of macroscopic, classical thermal machines is based on the paradigm that their dynamics is represented by an adiabatic sequence of equilibrium states wherein the system and the reservoir are separable \cite{schwabl2006statistical}. In recent years, it has turned out that in the quantum domain this paradigm does not commonly lead to heat-machine operation that is distinctly quantum mechanical \cite{benenti2017fundamental,binder2018thermodynamics}. To describe non-equilibrium thermal machines in the quantum domain, the leading approaches \cite{benenti2017fundamental,binder2018thermodynamics,klatzow2019experimental,kosloff2013quantum,kosloff2017quantum} have resorted to Born-Markov master equations, often under simplified assumptions: for example, individual strokes in an Otto cycle have been treated independently and then “glued” together to a full cycle. This in turn requires complete system-reservoir thermalization and separability in the appropriate strokes, thus remaining within the foregoing paradigm of classical thermodynamics.

A broader dynamical description of thermal machines that is fully reconciled with quantum mechanics should, however, account for the non-separability and correlations of the reservoirs with the system (the working medium - WM): these features are generic at low temperatures \cite{weiss12} and inevitably affect the machine operation \cite{gelbwaser2015thermodynamics}. Their treatment requires us to venture beyond the Born-Markov approximation \cite{breuer02}. Considerable progress in this direction has been recently achieved by revealing  non-Markovian and correlation effects in heat machines \cite{carrega2016energy,guarnieri2016energy,aurell2017work,pezzutto2019out,uzdin2016quantum,abiuso2019non}. In particular, it has been shown \cite{mukherjee2020anti} that non-Markovianity under fast driving of the WM can lead to a power boost of strictly quantum origin (“quantum advantage”) which is a manifestation of the anti-Zeno effect, whereby fast driving can enhance the system-reservoir coupling \cite{kofman2000acceleration,kofman2004unified,erez2008thermodynamic}. An important finding has been that switching on and off the WM-reservoir coupling in a stroke has  to  be fully accounted for,  as it may strongly influence the power output and efficiency  \cite{wiedmann2020non,wiedmann2021non}. Another development concerns the possibility of maintaining  coherence as a resource of the machine in either the WM \cite{uzdin2015equivalence} or the piston mode  \cite{ghosh2018two}.

On the methodological side, non-perturbative treatments of the non-equilibrium quantum dynamics have been established, such as generalized master-Floquet equations \cite{restrepo2016driven} and exact approaches based on the path integral representation of the reduced density operator of the WM \cite{weiss12}. Among the latter treatments one finds specifically the Hierarchical Equations of Motion (HEOM) \cite{tanimura89,kato2016quantum,Meng2021}, the Stochastic Liouville-von Neumann Equation(SLN) \cite{motz2018rectification, stockburger02}, the multilayer multiconfiguration time-dependent Hartree (ML-MCTDH) approach\cite{yang2020heat,velizhanin2008heat,wang03}, the nonequilibrium Green’s functions (NEGF) \cite{esposito2015quantum,esposito2015nature,carrega2022engineering}, and the continuous-time quantum Monte Carlo (CT-QMC) algorithm\cite{yamamoto2018heat,gull11} that have been recently applied  to explore quantum heat phenomena. These approaches provide us with powerful tools for exploring the mostly unfamiliar domain of quantum thermodynamics far from equilibrium, particularly in the presence of low-temperature mesoscopic or microscopic reservoirs. 

Current experimental studies of heat machines encompass a variety of quantum platforms: a single trapped ion as WM \cite{rossnagel2016single}; superconducting-circuit WM coupled to electronically engineered reservoirs \cite{cottet2017observing,pekola2015towards,ronzani2018tunable,senior2020heat,meschke2006single,pascal2011circuit}, nitrogen vacancy center WM coupled to nuclear spin baths in diamond \cite{klatzow2019experimental}, phonon reservoirs in solid-state circuits \cite{schwab2000measurement}, carbon nanotubes acting as mechanical resonators \cite{chang2006solid}; spin-wave or phonon-reservoirs in systems of trapped ions \cite{pruttivarasin2011trapped}, as well as inelastic spin-exchange collisions giving rise to quantum-controlled heat transfer directionality \cite{bouton2021quantum}. Although remarkable in terms of control and fabrication, these machines mostly operate in the domain of classical thermodynamics or only exhibit limited quantum effects. Yet experiment has thus far not implemented theoretical proposals for heat machines with advantageous properties of quantum origin \cite{benenti2017fundamental,binder2018thermodynamics,kosloff2014quantum,gelbwaser2015thermodynamics,wiedmann2020non,mukherjee2020anti,anders2013thermodynamics,Perarnau2018strong,lobejko2020thermodynamics,ghosh2017catalysis}. Particularly promising for experiment are theoretical predictions that non-overlapping spectra of thermal reservoirs allow for minimal designs of continuously operating quantum heat machines \cite{mukherjee2020anti,gelbwaser2013minimal,naseem2020minimal,gelbwaser2015work,gelbwaser2014heat,ghosh2018two}. Such continuous operation is experimentally advantageous as it does not require on-off-switching of the WM-reservoir thermal couplings and thus circumvents the conceptual difficulties  that are inherent in conventional stroke (reciprocal) heat machines, for example, a four stroke quantum Otto engine \cite{wiedmann2020non,wiedmann2021non,liu2021periodically}.

\begin{figure}
\centering
\includegraphics[width=12cm]{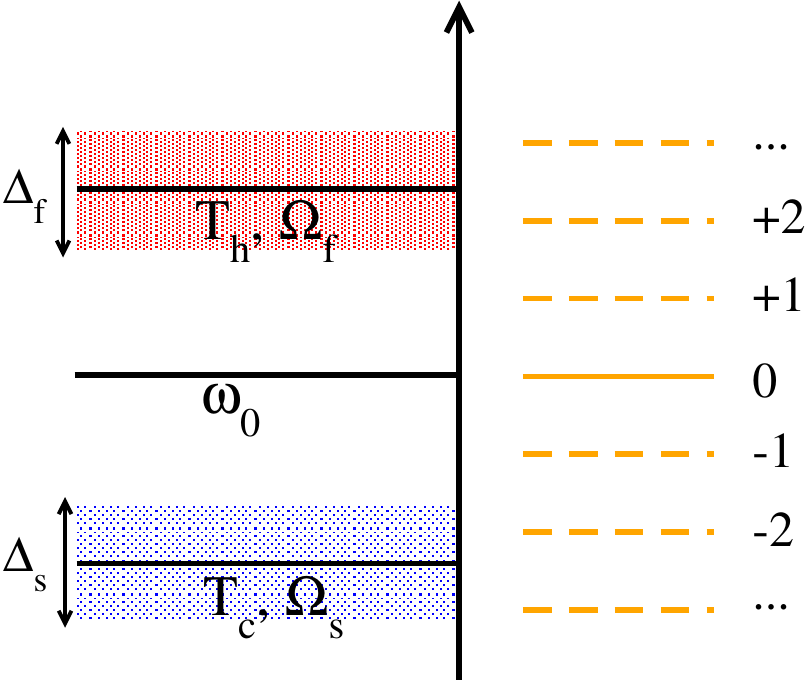}
\caption{A schematic view of the proposed quantum thermal machine in the frequency domain. The transition frequency of a two level system (TLS) with static level spacing $\omega_0$ is periodically  modulated as $\omega(t) = \omega_0 + \lambda \cos(\omega_s t)$ [see Eq. (\ref{Eq:wq})]. This modulation gives rise to side-bands $k\, \omega_s, k\in \mathcal{Z}$ (orange). In this way, the TLS realizes a working medium that interacts continuously with two reservoirs at different temperatures $T_h>T_c$ having localized spectral distributions at low frequencies around $\Omega_s$ (slow bath) and at high frequencies around $\Omega_f$ (fast bath) with respective bandwidths $\Delta_{\alpha}$.}
\label{fig1}
\end{figure}

Here we consider, as illustrated in Fig.~\ref{fig1}, a minimal design of a continuously operating heat machine: the WM is a two-level system (qubit) whose transition frequency is periodically modulated \cite{mukherjee2020anti,gelbwaser2013minimal,naseem2020minimal,gelbwaser2015thermodynamics,gelbwaser2014heat} such that it is in alternating spectral overlap with either the hot or the cold reservoir. The innovation  in the present design  is that the reservoir spectra are separated by a  bandgap and are therefore completely disjoint. Bandgap reservoirs arise in photonic crystals \cite{kofman1994spontaneous,lambropoulos2000fundamental} but can also be realized in superconducting circuits by placing transmission lines or LC-oscillators as interfaces between transmon qubits and ohmic resistors. 

To augment quantum effects, we adopt the strong system-reservoir coupling regime that arises near bandgaps \cite{kofman1994spontaneous}. We take the cold reservoir to be at zero temperature ($T_c = 0$). Under such conditions, we reveal highly nontrivial features such as anharmonicity of the WM reponse, non-equilibriation of the WM in finite-time cycles as well as quantum coherences and correlations of the WM-reservoir complex. Those features are revealed by the HEOM approach \cite{tanimura89,tanimura06,tanimura2020numerically,kato15,song17a}, a non-perturbative simulation technique of open quantum systems which is derived from the formally exact representation of the reduced density matrix in terms of path integrals \cite{weiss12,feynman63}. The HEOM has the additional benefit that, as part of the hierarchy, perturbative approaches such as the Redfield and the Lindblad master equations can be easily computed by the same code. In particular, the first order truncation of HEOM gives a generalized time-nonlocal master equation which allows us to find asymptotic Floquet states within a non-Markovian, perturbative method  \cite{magazzu2017asymptotic,magazzu2018asymptotic}. By comparing the results of these various approaches, insight is provided into the applicability of the perturbative treatments and the relevance of higher order quantum correlations. 

This paper is arranged as follows. In Sec. \ref{sec:model}, we present the heat engine model, whose operation essentially depends on the designed reservoir spectra. In Sec. \ref{sec:technique}, we outline the exact HEOM approach, its approximated treatment, and its use to extract heat currents. Simulation results are presented in Sec. \ref{sec:results}, where  HEOM data are shown together with the Redfield-plus approximation and a Markovian treatment. The analysis of sideband (Floquet) resonances follows in Sec.~\ref{sec:sidebands}. The power boost of the engine that reflects the non-Markovian anti-Zeno quantum advantage is discussed in Secs.~\ref{sec:nonMarkov1} and \ref{sec:nonMarkov2}. The conclusions are presented in Sec. \ref{sec:conclusion}.

\section{Minimal design of thermal machine with bandgap reservoirs}
\label{sec:model}
\begin{figure}
\centering
\includegraphics[width=12cm]{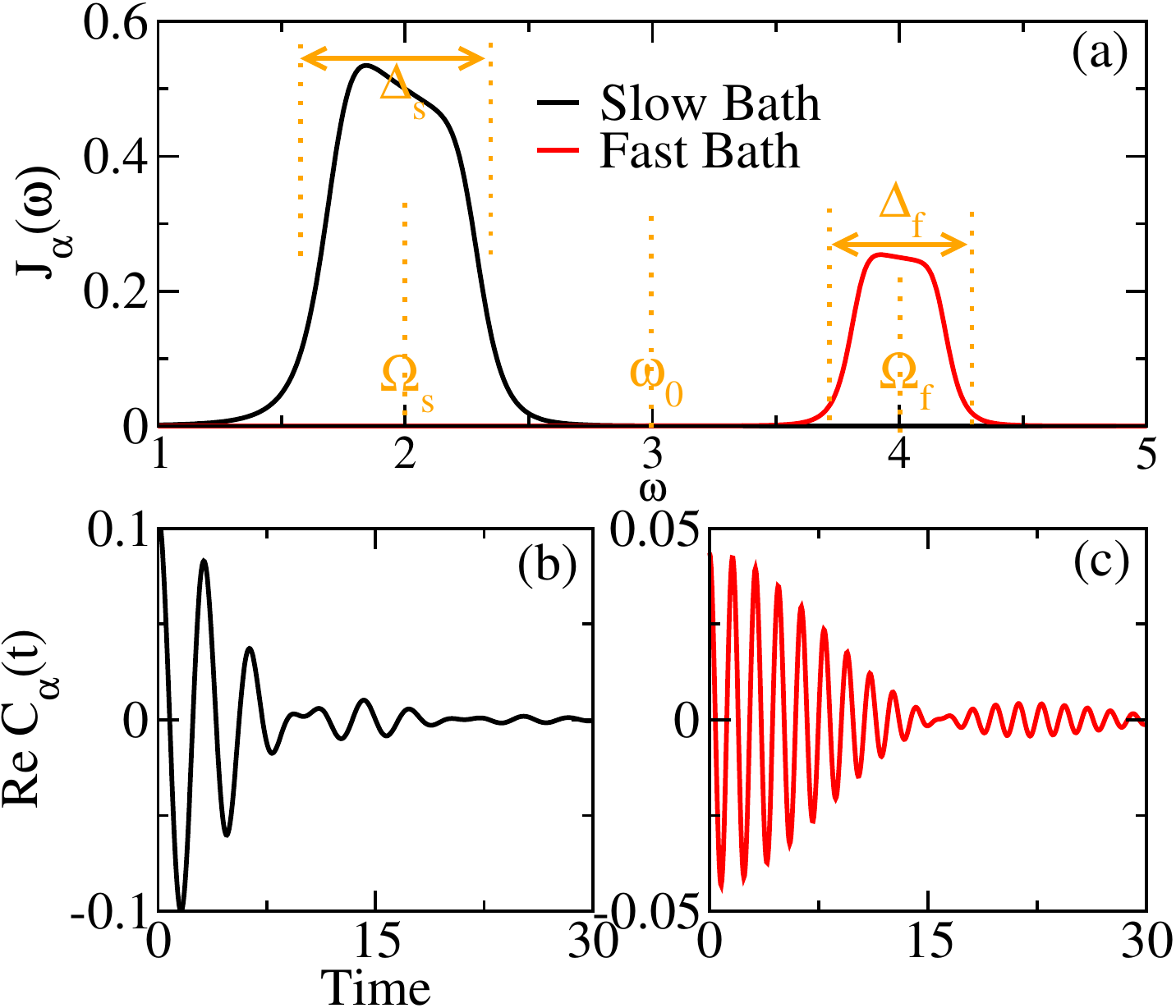}
\caption{Non-overlapping spectral densities sperated by a bandgap as in Fig.~\ref{fig1} for two thermal reservoirs interacting with a frequency modulated TLS with static transition frequency $\omega_0$ (a). The real parts of the corresponding correlation functions are shown $C_\alpha(t)$ for the slow (b) and the fast (c) bath. The parameters are: $\kappa_s = \kappa_f = 1.0$, ($\Omega_s$, $T_c$) = (2.0, 0.0), and ($\Omega_f$, $T_h$) = (4.0, 2.0).}
\label{fig2}
\end{figure}

Quantum heat engines (QHEs) can be theoretically studied within the framework of
open quantum dynamics\cite{kosloff2014quantum,gelbwaser2015thermodynamics,wiedmann2020non,grifoni1998driven,magazzu2018probing,traversa2013generalized,magazzu2018asymptotic,kosloff2014,gelbwaser2013minimal}.
Within this framework, the total Hamiltonian
consists of three parts,
\begin{equation}\label{Eq:ht}
  H(t) = H_0(t) + H_{I} + H_B \;\;,
\end{equation}
where $H_0(t)$ pertains to a system  that acts as WM subject to external driving
\cite{paauw2009tuning,ronzani2018tunable},
$H_B$ represents two thermal reservoirs (baths) at temperatures, 
$T_h > T_c$, henceforth denoted as hot reservoir and cold reservoir temperatures, respectively,
and $H_{I}$ is the coupling between the WM and the reservoirs. Here, we consider a paradigmatic WM, namely, a TLS with periodic frequency modulation, i.e.,
\begin{equation}\label{Eq:wq}
  H_0(t) = \hbar\omega(t)\sigma_{+}\sigma_{-}\;\;,
\end{equation}
with $\sigma_{\pm} = \frac{1}{2}\;(\sigma_x \pm i\sigma_y)$ written in terms of the Pauli matrices $\sigma_n,\;n = x,y,z$. The modulation of the transition frequency is taken to be of the form
\begin{equation}\label{Eq:wqt}
  \omega(t) = \omega_0 + \lambda \cos(\omega_s t)\;\;,
\end{equation}
whose amplitude $\lambda$ and driving frequency $\omega_s$ are the main variable parameters in the following analysis. We shall $\hbar = k_{B} = 1$. In the context of a thermal machine, the situation, where $\omega(t)<\omega_0$ corresponds to an expansion of the WM, while $\omega(t)>\omega_0$ represents its compression. Within the set-up illustrated in Fig.~\ref{fig2}, a conventional heat engine then operates such that the low (high) frequency bath has the lower (higher) temperature $T_c$ ($T_h$). However, we will also consider the reversed situation and demonstrate that it may induce even stronger heat fluxes. 

The bare system Hamiltonian is diagonal in the eigenbasis of $\sigma_+\sigma_-=(\sigma_z+1)/2$, i.e. $\sigma_+\sigma_- |q\rangle = q |q\rangle$ with $q = 0,\;1$. Consequently, the bare system time 
evolution operator $\mathcal{U}(t)= \mathcal{T}\exp[-i \int_0^t d\tau H_0(\tau)]$ can easily be expressed as 
\begin{equation}\label{Eq:quasi}
\begin{split}
\mathcal{U}(t) 
&= \exp\left\{  
    -i\int_0^td\tau\, [\omega_0 + \lambda\cos(\omega_s\tau)]{\sigma}_+{\sigma}_-
    \right\} \\
&= \sigma_- \sigma_+ +  \sigma_+\sigma_-\sum_{k=-\infty}^{+\infty}  J_k\left(\frac{\lambda}{\omega_s}\right)\  
{\rm e}^{-i\,\omega^{(k)}\, t}\;\;,
\end{split}
\end{equation}
where  $\omega^{(k)} = \omega_0 + k\omega_s$ are the quasi-energies (denoted by orange lines in 
Fig. \ref{fig1}) and $J_k(\cdot)$ are  Bessel functions of the first kind. 
The WM has thus multiple equidistant quasi-energy levels (in agreement with Floquet theory) which, as we will show, has prominent effects on quantum heat transport. For $\lambda/\omega_s\to 0$ one has $J_{|k|\neq 0}(\lambda/\omega_s)\to 0$, $J_0(\lambda/\omega_s)\to 1$ so that the system effectively reduces to a static system with level spacing $\omega_0$. 

As thermal reservoirs we consider quasi-continua of independent harmonic degrees of freedom with localized spectral distributions, namely, a slow bath $H_s$ composed of  oscillators in the frequency range below $\omega_0$ and a fast bath $H_f$ with oscillator frequencies above $\omega_0$. Accordingly, one can write
\begin{equation}\label{Eq:hbath}
\begin{split}
  H_B &=  H_s + H_f  \\
      &=
   \sum_{\alpha=s,f} \left\{ \sum_{i=1}^{\infty}
   \frac{p_{\alpha}^2}{2m_{\alpha,i}} + \frac{1}{2}m_{\alpha,i}\omega_{\alpha,i}^2 x_{\alpha,i}^2  \right\}\, ,
\end{split}
\end{equation}
where $m_{\alpha,i}$, $\omega_{\alpha,i}$, $x_{\alpha,i}$ and
$p_{\alpha,i}$ denote the mass, frequency, coordinate and
momentum of the $i$th oscillator in the $\alpha$th reservoir.
The bilinear interaction between these reservoirs and the WM is given by
\begin{equation}\label{Eq:hsb}
  H_{I} =  \sum_{\alpha=s,f} X_{\alpha} \sigma_x + \frac{1}{2} \sum_{\alpha=s,f} \mu_\alpha  \;\;,
\end{equation}
with $X_{\alpha} = \sum_{i}c_{\alpha,i}x_{\alpha,i}$ denoting a collective 
coordinate of the $\alpha$th reservoir and $c_{\alpha,i}$ being the coupling 
constant of the $i${\rm th} mode with $\alpha$. 
The last term guarantees that the reservoirs only act dynamically upon the system without any
coupling-induced distortion of the system. We note  that the total Hamiltonian in 
Eq. (\ref{Eq:ht}) can be transformed into the form of the conventional spin-boson model \cite{weiss12} 
by the canonical transformation $\tilde{H}(t) = SH(t)S^\dagger$ with   $S = \frac{1}{2}(\sigma_x +\sigma_z)$ which implies $\sigma_z\to \tau_x$ and $\sigma_x\to \tau_z$, where $\tau_k$ denote Pauli matrices in the rotated basis. 

The effect of the two reservoirs on the WM is completely described by the coupling weighted spectral densities summed over all modes
\begin{equation}
\label{Eq:dsp}
  J_{\alpha}(\omega) = \frac{\pi}{2}\sum_{i}
  \frac{c_{\alpha,i}^2}{m_\alpha\omega_{\alpha,i}}\delta(\omega - \omega_{\alpha,i}) \;\;
\end{equation}
which leads to $\mu_\alpha=(2/\pi) \int_0^\infty d\omega J_\alpha(\omega)/\omega$. According to our setting of two reservoirs with localized spectral distributions and only negligible overlap among them, the following continuum form of spectra is convenient, i.e.,
\begin{equation}\label{eq:sp}
  J_{\alpha}(\omega) = \frac{\kappa_{\alpha}\;\xi_\alpha^8\;\omega}
  {(\omega^2-\Omega_{\alpha}^2)^6 + \omega^2\xi_\alpha^{10}} \;\;, \ \alpha=s, f
\end{equation}
with central frequencies $\Omega_f>\omega_0>\Omega_s$  and frequency scales $\xi_\alpha$ which determine the widths $\Delta_\alpha$ of the spectral distributions. In what follows, all masses are set to unity and all frequencies are scaled with $\xi_s=\xi_f$ so that the dimensionless bandwidths $\Delta_\alpha$ are of order unity. The WM continuously interacts with the reservoirs with varying spectral overlap due to the external driving (see Figs. \ref{fig1} and \ref{fig2}). At resonance, the effective couplings read
\begin{equation}
    \kappa_{\alpha}^{\rm eff}
    = \frac{J_{\alpha}(\Omega_{\alpha})}{\Omega_{\alpha}}
    = \frac{\kappa_{\alpha}}{\Omega_{\alpha}^2} \;\;.
\end{equation}
As noted above, in order to focus on the quantum regime with strong non-Markovian effects, we consider the cold reservoir to have $T_c=0$. This immediately implies that  our machine cannot be  a quantum refrigerator. 

\section{Simulation Techniques}\label{sec:technique}
The described setting is highly non-trivial, first, because we consider localized, non-overlapping reservoir (bath) spectra and the machine is operated at low temperatures. Hence, correlations between WM and the thermal reservoirs are expected to be so strong that conventional master equations are not applicable. We thus rely on the exact quantum dynamical simulations within the Hierarchical Equations of Motion (HEOM) approach.

\subsection{Hierarchical equations of motion}
\label{sec:heom}

Here, we briefly describe the essence of the HEOM approach and its derivation from the path integral expression of the reduced density matrix. For the sake of simplicity, we consider only a single reservoir for a Hamiltonian of the form (\ref{Eq:hbath}) and assume
factorized initial states at the time zero $t=0$,
$\rho_T(0) = \rho_m(0)\otimes e^{-\beta H_B}/Z_B$, where $Z_B = \rm{Tr}\, e^{-\beta H_B}$ and $\rho_m$ is the density operator of the relevant system (WM). The generalization to correlated initial states has also been discussed \cite{tanimura14,song15b}.

In path integral representation \cite{feynman63} the reduced density operator is obtained as
\begin{equation}\label{Eq:rdo}
\begin{split}
  \rho_m(t)=&\int\mathcal{D}q^{+}(t)\mathcal{D}q^{-}(t) e^{i\{S_{+}[q^{+}(t)]-S_{-}[q^{-}(t)]\}}
  \mathcal{F}[\sigma_x^{+}(t),\sigma_x^{-}(t)]\rho_m(0) \;\;.
  \end{split}
\end{equation}
 Generally, a continuous system coordinate can be discretized using a
 system-specific discrete variable representation (DVR) \cite{echave92}.
 Within the HEOM and for the TLS-WM considered here, a representation in terms of the eigenstates
  $|q\rangle \in \{|0\rangle,|1\rangle\}$ is
 convenient. The coordinates $q^{+}(t)$ and $q^{-}(t)$ denote forward and backward system
 paths, respectively, and $S_{\pm}[q^{\pm}(t)]$ the corresponding actions,
\begin{equation}
    S_{\pm}[q^{\pm}(t)] = -\int_0^t d\tau \omega(\tau) q^{\pm}(\tau)\;\;.
\end{equation}
These paths $q^\pm(t)$ directly determine also the $\sigma_x^{\pm}(t)$ according to 
\begin{equation}
    \sigma_x^{\pm}(t) = \langle q^{\pm}(t^+)| {\sigma}_x|q^{\pm}(t)\rangle \;\;, 
\end{equation}
where $t^+$ denotes the time slice on the forward and backward
paths that follows the time slice $t$. 

The effective impact of the reservoir onto the system dynamics is described by the influence functional \cite{feynman63} which reads
\begin{equation}\label{Eq:fv-if}
\begin{split}
  \mathcal{F}[\sigma_x^{+}(t),\sigma_x^{-}(t)] =
   \exp&\left\{ -\int_{0}^{t}ds [\sigma_x^{+}(s)-\sigma_x^{-}(s)] \right. \\
&\times \left.\int_{0}^{s}d\tau
  \left[C(s-\tau)\sigma_x^{+}(\tau) - C^{\ast}(s-\tau)\sigma_x^{-}(\tau) \right]
  \right\}\;\;.
\end{split}
\end{equation}

The derivation of the real-time HEOM starts by first
expanding \cite{tanimura89,jin08} or fitting \cite{tang15,meier99} the bath 
correlation function as a sum 
of exponential terms, i.e.
\begin{equation}\label{Eq:cdcp}
\begin{split}
  C(t) &= \frac{1}{Z_B}{\rm Tr}_B\left[e^{-\beta H_B} X(t)X(0) \right] \\
  &= \int_{-\infty}^{+\infty}\frac{d\omega}{\pi} J(\omega)\, n_\beta(\omega)e^{-i\omega t}\\ &
  =\sum_{k} d_k e^{-\gamma_k t}\;\; \ \ \ \rm{for}\; t>0\;\;
\end{split}
\end{equation}
with the collective bath operator $X$ as in Eq. (\ref{Eq:hsb}) and the Bose-Einstein distribution $n_\beta(\omega) = 1/(1-e^{-\beta\omega})$. Accordingly, the spectral function $S(\omega)=J(\omega)n_\beta(\omega)$ obeys $S(\omega)=S(-\omega)+J(\omega)$ with $J(-\omega)=-J(\omega)$ as in Eq. (\ref{eq:sp}).
The $d_k$ in (\ref{Eq:cdcp}) denote proper coefficients in an expansion in terms of exponentials with proper coefficients $\gamma_k$. All characteristics of the reservoirs are expressed  by the correlation functions which are depicted in Fig.~\ref{fig2} for the slow and fast reservoirs in Eq. (\ref{eq:sp}), respectively. In contrast to conventional Ohmic-type spectral distributions $J(\omega)\propto \omega$ that at higher temperatures lead to an exponential decay of the auto-correlation $C(t)$, we here observe long correlation times $\tau_R\gtrsim 30$ (in arbitrary units) that exceed the driving (modulation) periods $\tau_s=2\pi/\omega_s$ if $\omega_s\gtrsim 2\pi/\tau_R\approx 0.2$. Consequently, a separation of time scales on which Markovian perturbative approaches are based, does not exist and memory effects are strong as we will see below. The limited bandwidth of the reservoirs yields beating patterns, particularly for the high frequency bath.

By resorting to the following auxiliary density operator (ADO) definition,
\begin{subequations}\label{eq:ado1}
\begin{equation}
\begin{split}
\rho_{\bf n}(t)
=&\int \mathcal{D}q^+(t)\mathcal{D}q^-(t)
e^{i\{S_{+}[q^+(t)]-S_{-}[q^-(t)]\}} 
\prod_{k} [\phi_k(t)]^{n_k}
\mathcal{F}[\sigma_x^+(t),\sigma_x^-(t)] \rho_m(0)\;\;;
\end{split}
\end{equation}
\begin{equation}
\begin{split}
\phi_k(t) =  -i\int_0^t ds
\Big[\sigma_x^+(s)d_k e^{-\gamma_k(t-s)} -\sigma_x^-(s) d_k^{*} e^{-\gamma_k(t-s)} \Big]\;\;,
\end{split}
\end{equation}
\end{subequations}
the  HEOM formulation leads to the equation of motion \cite{tanimura89,ishizaki05,tanimura06,xu17,zhang2020hierarchical,ikeda2020generalization,yan2020new,tanimura2020numerically}
\begin{align}
\label{eq:real}
\frac{d\rho_{\bm n}(t)}{dt}=&-\left(i\mathcal{L}_0(t)
+\sum_{k}n_{k}\gamma_{k}\right){\rho}_{\bm{n}}(t)-
i\left[{\sigma}_x,\sum_{k}
{\rho}_{{\bm{n}}_{k}^{+}}(t)\right]\nonumber\\
&-i\sum_{k}n_{k}\left(d_{k} {\sigma}_x
{\rho}_{{\bm{n}}_{k}^{-}}(t)-d_{k}^{*}{\rho}_{{\bm{n}}_{k}^{-}}(t){\sigma}_x\right) \;\;,
\end{align}
which shows that the WM interacts via ${\sigma}_x$ with the collective bath force.
The ADOs
$\rho_{\bf{n}}$s are labeled by  the subscript $\bf{n}$ denoting a set of integers
$\{n_{1},...,n_{k},...\}$, with $n_{k} \geq 0$ associated with
the $k$th exponential term in Eq. (\ref{Eq:cdcp}); ${\bf{n}}_{k}^{+}$ and ${\bf{n}}_{k}^{-}$ denote
$\{n_{1},...,n_{k}+1,...\}$, and $\{n_{1},...,n_{k}-1,...\}$, respectively.
The super-operator acting on these ADO is defined as $\mathcal{L}_0(t){\rho_{\bm{n}}}=\left[H_{0}(t), \rho_{\bm{n}}\right]$. The WM reduced density operator in this notation is $\rho_{m}=\rho_{\{0,...0,...\}}$.

Assuming that $\rho_{\bm 0}(t)$ is of order one, the magnitude of
$\rho_{\bm n}(t)$ is proportional to $\prod_k d_k^{n_k}$, which may be divergent for strong system-bath coupling $\eta$ as $|{\bm n}|\doteq n_1 + n_2 + ....+n_k+...$ increases. Therefore, the original HEOM \cite{tanimura89,tanimura90} is re-scaled and combined with on-the-fly filtering methods \cite{shi09b} to solve this problem efficiently.
In our simulations, we choose the following rescaling,
\begin{equation}
    \tilde{\rho}_{\bm n}(t) =
    \left(\prod_k n_k! \; |d_k|^{n_k}\right)^{-1/2}\rho_{\bm n}(t)\;\;,
\end{equation}
so that Eq. (\ref{eq:real}) is recast as
\begin{small}
\begin{equation}\label{eq:scaled}
\begin{split}
\frac{d \tilde{\rho}_{\bm{n}}(t)}{dt}=&-\left(i\mathcal{L}_0(t)
+\sum_{k}n_{k}\gamma_{k}\right) \tilde{\rho}_{\bm{n}}(t) 
-i\sum_k \sqrt{(n_k+1)|d_k|}\left[{\sigma}_x,
\tilde{\rho}_{{\bm{n}}_{k}^{+}}(t)\right] \\
&-i\sum_{k}\sqrt{\frac{n_k}{|d_k|}}\left(d_{k} {\sigma}_x
\tilde{\rho}_{{\bm{n}}_{k}^{-}}(t)-d_{k}^{*}\tilde{\rho}_{{\bm{n}}_{k}^{-}}(t){\sigma}_x\right) \;\;,
\end{split}
\end{equation}
\end{small}
The magnitude of $\tilde{\rho}_{\bm n}(t)$ is proportional to $\prod_k \sqrt{|d_k|^{n_k}/n_k!}$ and decays to zero
for high hierarchical levels. Therefore, we can set $\rho_{\bm n}(t) = 0$ if $|\rho_{\bm n}^{{\rm max}}(t)|<\delta$, where $\delta$ denotes the error tolerance ( here we set $\delta=10^{-7}$). More advanced algorithms to support the efficiency and numerical stability can be found in Refs.\ \cite{cui2019highly,shi2018efficient,borrelli2019density,dunn2019removing,yan2020new,ikeda2020generalization,zhang2020hierarchical,tanimura2020numerically,yan2021efficient}.

According to (\ref{eq:scaled}) we expect the reduced density to approach a periodic steady state $\rho_m(t)\to \rho^{(st)}_m(t)= \rho^{(st)}_m(t+\tau_s)$ with $\tau_s = 2\pi/\omega_s$ being the driving (modulation) period. This periodicity characterizes all single time-dependent observables (one-point correlations) such as the excited (ground) state population $P_e(t)=\langle \sigma_+\sigma_-\rangle_t$ ($P_g=1-P_e$) or the heat currents $I_\alpha(t)$.  

\subsection{Perturbative treatment}
\label{subsec:sec-some}

Approximate treatments of open system dynamics have been developed to second order in the system-reservoir coupling. Together with a time scale separation between fast decaying reservoir correlations and relaxation dynamics of the reduced density operator, this leads to the Redfield master equation \cite{breuer02}. Interestingly, an extended Redfield equation is also obtained from the HEOM if it is curtailed at first-order, i.e.\ resricted to ADOs with $\sum_kn_k = 1$. Namely, in the interaction picture this Redfield equation has the form 
\begin{subequations}
\begin{equation}\label{Eq:drhos}
  \frac{d}{dt}\rho_m^{I}(t)
  = -i\sum_k[{\sigma}_x^{I}(t),\rho_{{\bf 0}_k^{+}}^{I}(t)] \;\;,
\end{equation}
\begin{equation}\label{Eq:drho1}
\frac{d}{dt}\rho_{{\bf 0}_k^{+}}^{I}(t)
= -\gamma_k\rho_{{\bf 0}_k^{+}}^{I}(t)
 -i[d_k{\sigma}_x^{I}(t)\rho_m^I(t) - d_k^{*}\rho_m^I(t)\sigma_x^{I}(t)] \;\;,
\end{equation}
\end{subequations}
where ${\sigma}_x^I(t)$, $\rho_s^{I}(t)$ and $\rho_{\bm{0}_k^{+}}^{I}(t)$ denote the system (WM), reduced density matrix and first-order ADOs in the interaction picture,
respectively. Note that Eqs. (\ref{Eq:drhos}) and (\ref{Eq:drho1}) have the same 
structure as their counterparts in the generalized Floquet theory \cite{traversa2013generalized,magazzu2017asymptotic,magazzu2018asymptotic}. Upon solving  Eq. (\ref{Eq:drho1}) and inserting it into Eq. (\ref{Eq:drhos}), one has
\begin{equation}
\label{Eq:heom2rd}
\begin{split}
\frac{d}{dt}\rho^{I}_{m}(t)
=& -\sum_{k}\int_0^{t}d\tau
e^{-\gamma_k(t-\tau)}
{\bf[} \sigma_x^I(t),
d_k{\sigma}_x^I(\tau)\rho_m^I(\tau)- d_k^{*}\rho_m^{I}(\tau){\sigma}_x^I(\tau){\bf]}  \\
=&-\int_{0}^{t}d\tau
[{\sigma}_x^{I}(t),C(t-\tau){\sigma}_x^{I}(\tau)\rho_m^{I}(\tau)-C^{*}(t-\tau)\rho_m^I(\tau){\sigma}_x^I(\tau)] \\
=&-\int_0^t d\tau {\rm Tr_B}\{
[H_{I}^I(t),[H_{I}^I(\tau),\rho_T^I(\tau)]]\}\;\;.
\end{split}
\end{equation}
In the above expression, the correlation function in Eq. (\ref{Eq:cdcp}) and the Born approximation  \cite{breuer02} have been used but {\em not} the Markov approximation. Thus, Eq. (\ref{Eq:heom2rd}) is an integro-differential equation nonlocal im time which is henceforth denoted {\em Redfield-plus} (Redfield$^+$) to distinguish it from the conventional Redfield formulation. It can be conventiently solved with the help of auxiliary variables \cite{frishman96,meier99,thanopulos08}.

In the  Markov limit one sets in Eq. (\ref{Eq:heom2rd}) $\rho_m^I(\tau)\to \rho_m^I(t)$, leading to the conventional time-local Redfield master equation
\begin{equation}\label{Eq:redfield}
\frac{d}{dt}\rho^{I}_{m}(t)
=-\int_0^t d\tau {\rm Tr_B}\{
[H_{I}^I(t),[H_{I}^I(\tau),\rho_T^I(t)]]\}\;\;.
\end{equation}
This equation can be easily solved in the time domain with the help of Eq. (\ref{Eq:cdcp}) through the use of the auxiliary operator
\begin{equation}
    q_k^I(t) = -i\int_0^td\tau\;d_k e^{-\gamma_k\;(t-\tau)} \sigma_x^{I}(\tau) \;\;.
\end{equation}
This substitution transforms Eq. (\ref{Eq:redfield}) into 
\begin{subequations}
\begin{equation}
    \frac{d}{dt} \rho_m^{I}(t) = -i\sum_k\; [\sigma_x^I(t),\;q_k^I(t)\rho_m^I(t) 
                                             + \rho_m^I(t)q_k^{I*}(t)] \;\;;
\end{equation}
\begin{equation}
    \frac{d}{dt}q_k^I(t) = -\gamma_k q_k^I(t) - i d_k \sigma_x^I(t) \;\;.
\end{equation}
\end{subequations}
The HEOM approach 
can been seen as an infinite-order extension of the Redfield-plus/Redfield approximation
\cite{xu17,xu2018convergence,trushechkin2019higher}. This
allows us to reveal consistently the impact of higher order system-reservoir correlations which
are particularly subtle for heat currents.

\subsection{Heat current, power, and efficiency}
In the framework of the HEOM, effects of the environment on the system dynamics can be obtained from the ADOs \cite{song17a,zhu12,kato15,kato2016quantum,duan2020unusual}. Here, we concentrate on the quantum heat current which is linear in the collective bath force $X$.
One starts, in the interaction picture, with the following two equations
\begin{subequations}
\begin{equation}
  \frac{d}{d t}\rho_m^I(t) = -i\sum_{k}[{\sigma}_x^I(t),\rho_{\bm{0}_{k}^{+}}^{I}(t)] \;\;;
\end{equation}
\begin{equation}
\begin{split}
\frac{d}{d t}\rho_m^I(t)
= &-i{\rm Tr}_{B}\{[{\sigma}_x^I(t)X^I(t),\rho_T^{I}(t)] \}
= -i[{\sigma}_x^{I}(t),{\rm Tr}_{B}\{X^I(t)\rho_T^I(t)\}] \;\; ,
\end{split}
\end{equation}
\end{subequations}
where $\rho_{T}^{I}(t)$ denotes
the total density matrix in the interaction picture. In the next step,  relations between first-order ADOs ($\rho_{\bm{0}_{k}^{+}}$)
in the HEOM and first-order moments of $X$ are constructed according to
\begin{equation}\label{}
  \sum_{k}\rho_{\bm{0}_k^{+}}^{I}(t) = \rm{Tr}_{B}\{X^I(t)\rho_T^I(t)\}\;\;.
\end{equation}
Higher-order relations can be found in Refs.\ \cite{song17a,zhu12}. 

In the presence of two thermal reservoirs, the above relation is inserted into the definition
\cite{esposito2015nature,song17a} derived from first principal \cite{kosloff2013quantum,kosloff2014quantum} to obtain the quantum heat current between WM and reservoirs, i.e.
\begin{equation}\label{Eq:qhc}
\begin{split}
I_{\nu}(t)
  &\equiv -\frac{d}{dt}\langle H_{\nu}+H_{{I}, \nu}\rangle
 = -i\langle [ H_{0}(t),{\sigma}_x X_{\nu}]\rangle \\
 &=-i{\rm Tr}_m\left\{[H_0(t),{\sigma}_x]
 {\rm Tr}_{B}\{X_{\nu}\rho_T\} \right\} \\
 &= \omega(t) \sum_{k}{\rm Tr}_m\left\{{\sigma}_y \rho_{\bm{0}_{k}^{+}}(t)\right\} 
 \;\;, \;\; k\in \nu {\rm th\; bath} \;\;,
\end{split}
\end{equation}
where $\nu=c, h$ indicates cold and hot bath (either slow or fast), respectively.  By definition, a positive value of $I_{\nu}(t)$ corresponds to energy flowing into system, while a negative value corresponds to its inverse. We note in passing that an alternative definition of heat current as minus the energy change in the reservoir, which is directly related to the system energy changes when the system-reservoir coupling energy is negligible. However, the situation is very different when studying strong coupling setups where the system is periodically driven and in this case changes in the coupling energy must be accounted for \cite{esposito2015nature,esposito2015quantum,liu2021periodically}.

Deeper insight into the operation of the thermal machine in the quantum regime is given by the normalized correlations between the slow and the fast reservoir modes mediated by the driven two level WM which can be easily obtained from the HEOM, i.e.\ 
\begin{equation}\label{Eq:xsxf}
    \langle X^I_sX^I_f\rangle_t = \frac{1}{C_s(0)C_f(0)}\sum_{k, l} {\rm Tr}_m 
       \{\rho^I_{\bm{0}_k^+,\bm{0}_l^+}(t)\} \;\;,
\end{equation}
with $k,l$ denoting slow and fast baths effective modes, respectively. These correlations are not present in the classical treatment nor in the standard Redfield master equation. 

In steady state, we have $\rho_{\bm{n}}^{sst}(t) = \rho_{\bm{n}}^{sst}(t + \tau_s)$, and also $I_{\nu}^{sst}(t) = I_{\nu}^{sst}(t + \tau_s)$. It is thus convenient to define 
\begin{equation}
    \bar{I}_{\nu} =\lim_{t \to \infty} \frac{1}{\tau_s}\int_t^{t+\tau_s}d\tau I_{\nu}^{sst}(\tau) \;\;
\end{equation}
as an average of the heat current over one driving period. When representing the current as $I_\nu^{\rm st}(t)=\sum_m \mathcal{I}_{\nu,m} \exp(-i m \omega_s t)$, this then implies $\bar{I}_{\nu}=\mathcal{I}_{\nu, 0}$.

The operation of the thermal machine as a heat engine is characterized by its power output and its efficiency. In steady-state, the power can be calculated directly from $\bar{I}_c$ and 
$\bar{I}_h$ according to 
\begin{equation}
\begin{split}
    P &= \lim_{t\to\infty} \frac{1}{\tau_s}\int_t^{t+\tau_s}d\tau \langle -\frac{\partial H_s(\tau)}{\partial \tau}\rangle 
    = \lim_{t\to\infty} \frac{1}{\tau_s} \int_t^{t+\tau_s}d\tau\; \omega(\tau)\; \dot{P}_e(\tau) \\
    &=\lim_{t\to\infty} \frac{i}{\tau_s} \sum_{k}\int_t^{t+\tau_s}d\tau\; \omega(\tau) \; 
     [\rho^{12}_{\bm{0}_{k}^{+}}(\tau) - \rho^{21}_{\bm{0}_{k}^{+}}(\tau) ] \\
    &= \lim_{t\to\infty} \frac{1}{\tau_s}\int_{t}^{t+\tau_s}d\tau [I_c(\tau) + I_h(\tau)] 
    = \bar{I}_c + \bar{I}_h\;\;.
\end{split}
\end{equation}
Here, $\rho^{12}_{\bf{0}_{k}^{+}}$ and $\rho^{21}_{\bf{0}_{k}^{+}}$ denote 
ADO elements and $P_e(t)=\langle \sigma_+\sigma_-\rangle_t$ denotes the excited state population.
From the above definition positive values of $P$ mean that heat is converted into work (heat engine operation), while negative $P$
means that work is dissipated in the reservoirs (dissipator). In steady-state, a formal definition of the efficiency is given by  
\begin{equation}
    \eta =\frac{P}{\bar{I}_h} 
           = 1 + \frac{\bar{I}_c}{\bar{I}_h} \;\;.
\end{equation}
Note that here the efficiency has a physical meaning only for positive values and is then a figure of merit for the heat engine. In addition, when $T_c = 0$, the efficiency is limited to $1$ by the first law of thermodynamics, i.e. energy conservation, which forbids an output that exceeds the input \cite{gelbwaser2015thermodynamics}, otherwise, is not.
With the above set of equations at hand, we are in a position to explore heat transfer properties of the quantum heat engine in more detail.

\section{Simulation results}
\label{sec:results}
In this section, we present numerical results based on the HEOM and  the approximate treatments Redfield$^+$ and Born-Markov Redfield, respectively. The total initial density matrix is taken to be a factorized state i.e.\ 
$\rho_T(0) = \rho_m(0)\otimes \prod_{\alpha} e^{-\beta_{\alpha} H_{\alpha}}/{\rm Tr}[e^{-\beta_{\alpha}H_{\alpha}}]$, whose time evolution is followed until a steady state is approached. In this regime, observables are calculated for various parameter sets, where thermal reservoirs are characterized by their temperature and central frequency  $(\Omega_\alpha, T_i)$, $\alpha=s, f$ and $i=h, c$ while bandwidths $\Delta_\alpha$ are kept constant throughout.  

As a first step, we compare data obtained from the approximate treatment with exact ones from the HEOM. The task is to find parameter domains for which the Redfield$^+$ is sufficiently accurate and conversely to identify, where it fails due to very strong WM-reservoirs correlations (strong non-Markovianity). 

In the numerical simulations, the spectral  function 
$S(\omega) = J(\omega)n_{\beta}(\omega)$ is properly fitted to an optimized rational function 
with tolerance $\delta_S \le 10^{-6}$ and then Fourier transformed to correlation functions
$C(t)$ written as a sum of exponential terms. Concurrently, and on-the-fly filtering \cite{shi09b}
algorithm is adopted in order to achieve high efficiency. Atomic units (a.u.) 
are used here in order to treat a variety of regimes.

\subsection{Perturbative versus exact treatment}
\label{subsec:hvsr}

\begin{figure}
\centering
\includegraphics[width=12cm]{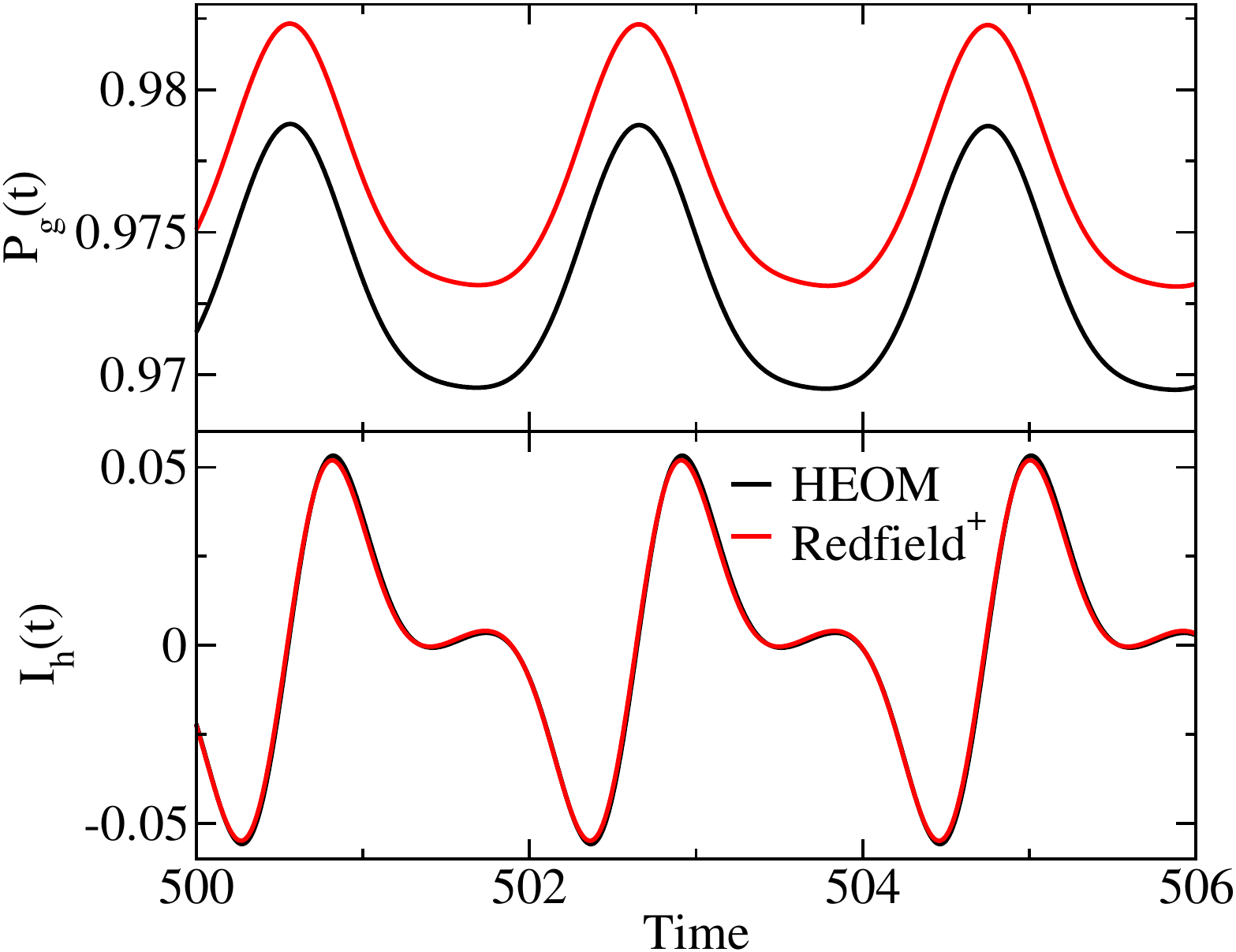}
\caption{Time-dependent dynamics of the ground-state population of the WM (top) and of the heat current $I_h(t)$ (bottom) in periodic asymptotic (steady) state simulated by HEOM and Redfield$^+$. The simulation parameters are (in a.u.): $\omega_0 = 3.0$, $\lambda = 1.0$, $\kappa_{\alpha} = 1.0$, $\omega_s$ = 3.0, ($\Omega_{s}$, $T_c$) = (2.0, 0.0), and 
($\Omega_{f}$, $T_h$) = (4.0, 2.0).}
\label{fig3}
\end{figure}

\begin{figure}
\centering
\includegraphics[width=12cm]{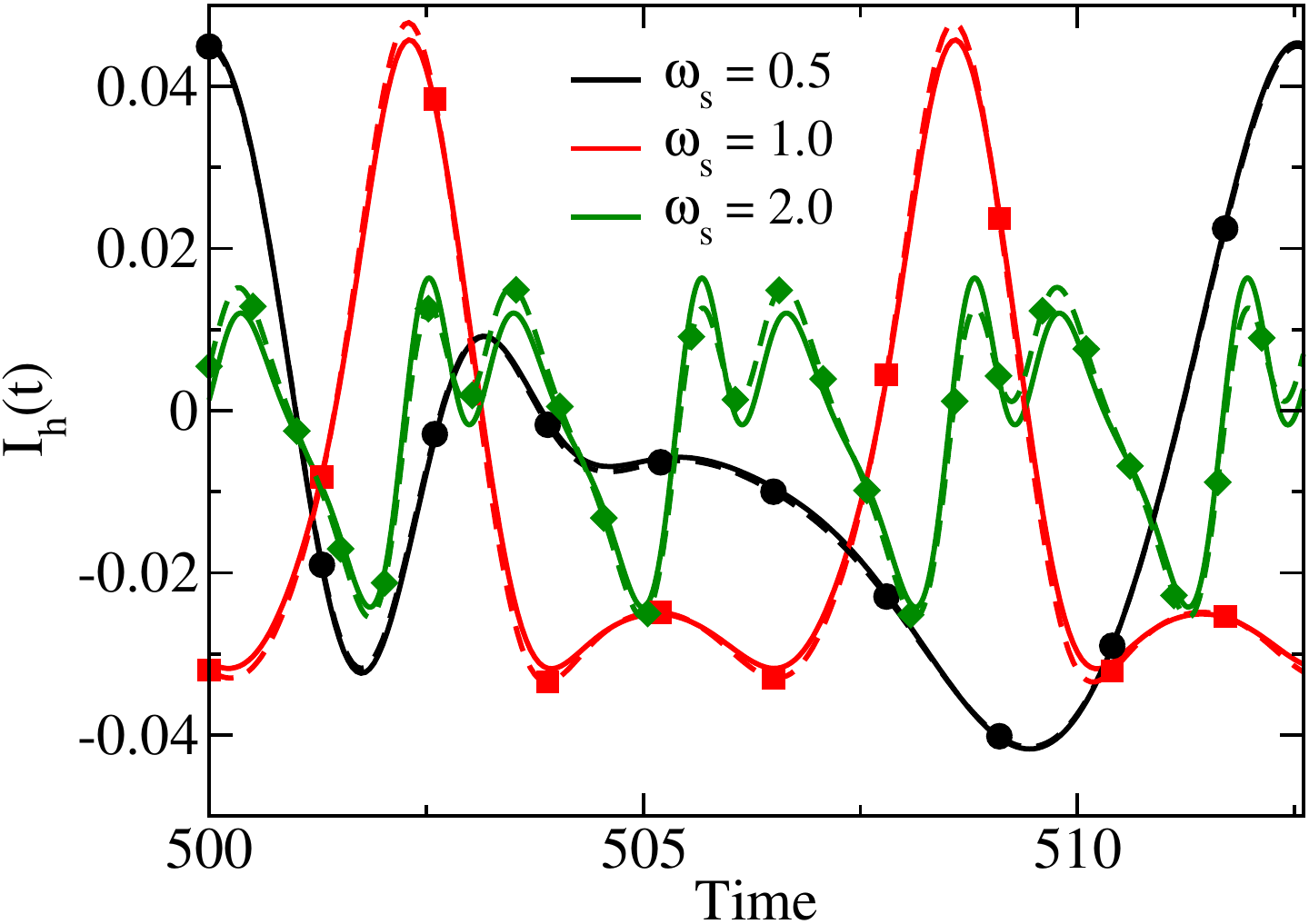}
\caption{Heat current dynamics $I_h(t)$ in periodic steady state simulated by HEOM (dashed lines) and Redfield$^+$ (solid lines) for various driving frequencies. The simulation parameters are (in a.u.): $\omega_0 = 3.0$, $\lambda = 1.0$, $\kappa_{\alpha} = 1.0$, ($\Omega_{s}$, $T_c$) = (2.0, 0.0), and ($\Omega_{f}$, $T_h$) = (4.0, 2.0).}
\label{fig4}
\end{figure}

Let us first recall the relevant time scales in the periodic steady state, i.e.\ the external modulation period $\tau_s=2\pi/\omega_s$ and the typical correlation time (memory time) of the thermal reservoirs $\tau_R$. At low temperatures $\tau_R \sim \beta$ so that Markovian treatments fail and particularly
 $\tau_s\ll \tau_R$. For bandgap environments additional time scales come into play, namely, the central band frequencies $\Omega_\alpha$ and respective widths $\Delta_\alpha$. 

\begin{figure}
\centering
\includegraphics[width=12cm]{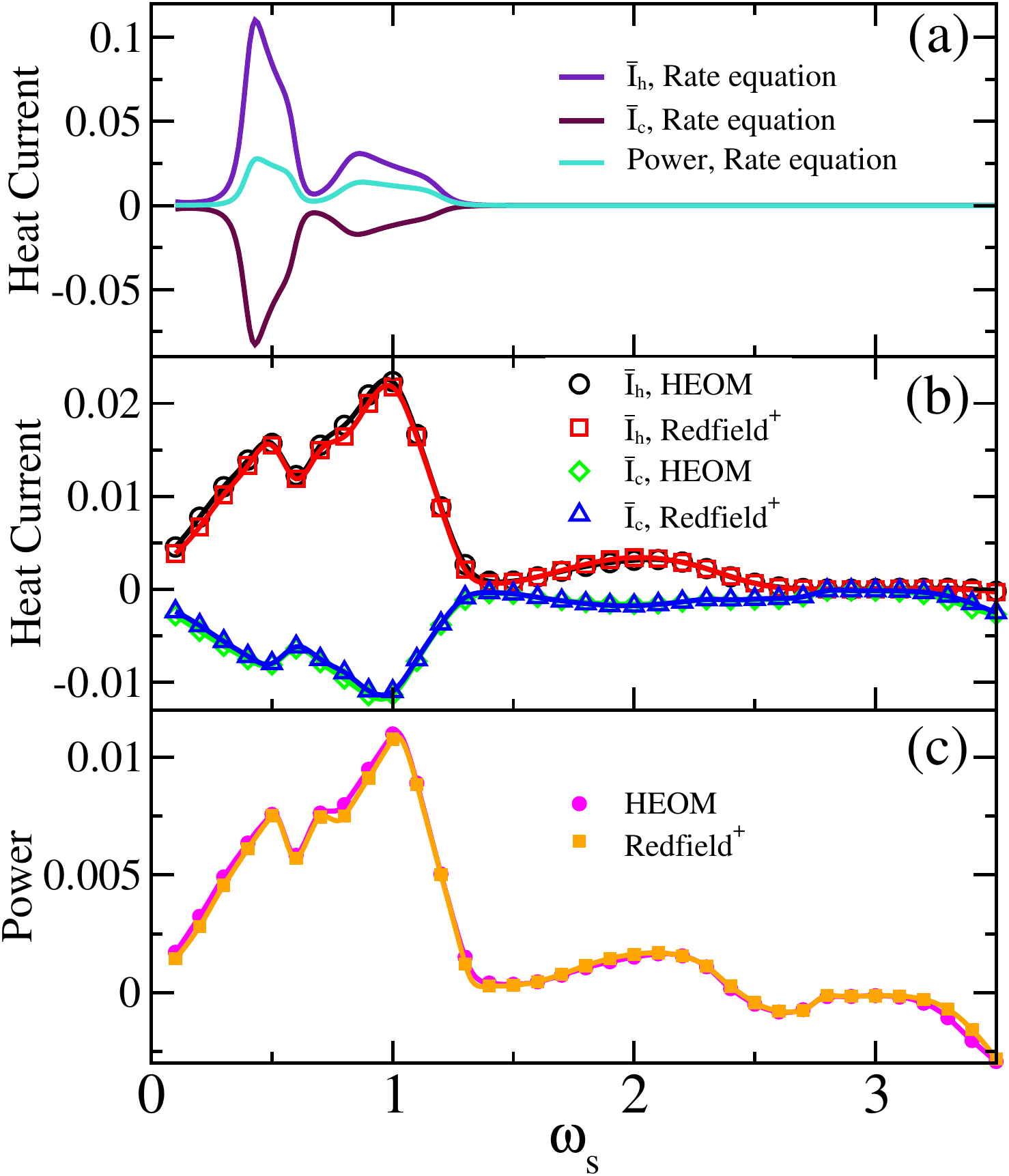}
\caption{Mean heat current versus driving frequency in the Born-Markov approximation (a) and according to the methods of Redfield$^+$ and HEOM (b). The net power corresponding to (b) is depicted (c).
The simulation parameters are the same as in Fig.~\ref{fig4}.}
\label{fig5}
\end{figure}

\begin{figure}
\centering
\includegraphics[width=12cm]{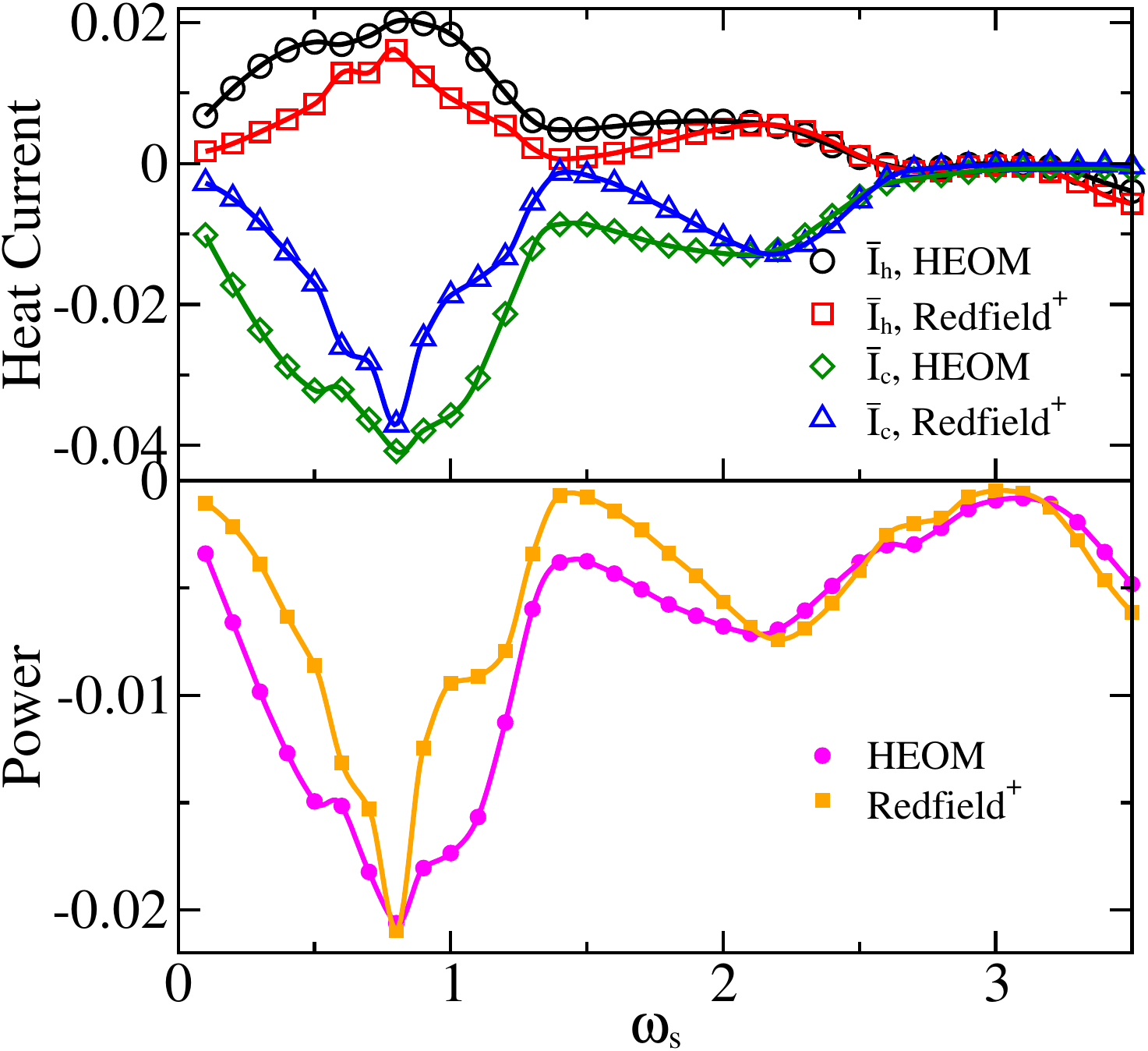}
\caption{Mean heat currents $\bar{I}_h$ and $\bar{I}_c$ according to Redfield$^+$ and HEOM (top) and the corresponding net power with the temperature gradient reversed compared to Fig.~\ref{fig5}, i.e.\ the slow bath is hot, the fast bath is cold.
The simulation parameters are (in a.u.): $\omega_0 = 3.0$, 
$\lambda = 1.0$, $\kappa_{\alpha} = 1.0$, ($\Omega_{s}$, $T_h$) = (2.0, 2.0), and 
($\Omega_{f}$, $T_c$) = (4.0, 0.0).}
\label{fig6}
\end{figure}

\begin{figure}
\centering
\includegraphics[width=12cm]{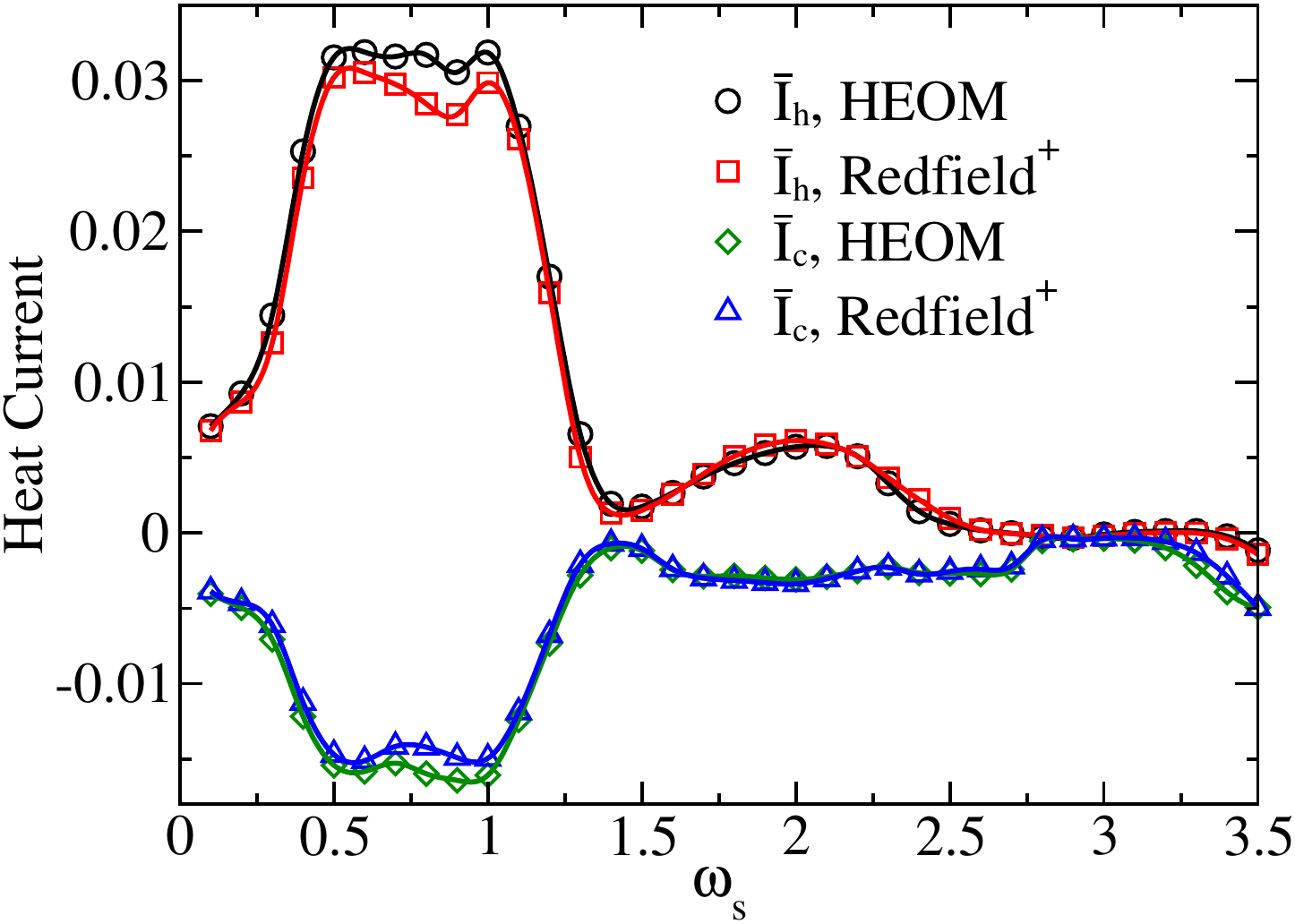}
\caption{Same as in Fig.~\ref{fig5} but with larger modulation amplitude $\lambda=1.5$.}
\label{fig7}
\end{figure}

Now, in Figs. \ref{fig3} to \ref{fig7} we compare  the performance of the perturbative Redfield$^+$ with the exact HEOM for various observables. The general outcome of this comparative analysis is that we can identify parameter regimes, where the approximate treatment provides quantitatively excellent results at least for heat currents. While Redfield$^+$ accounts to some extent for memory effects in the thermal reservoirs, it does so for sufficiently weak system-bath interaction. Consequently, the dynamics of the system operator in Eq. (\ref{Eq:heom2rd}) $\sigma_x(t)$ reflects the bare dynamics, i.e.,
\begin{equation}\label{eq:sigmax}
\begin{split}
\sigma_x(t) 
&= \mathcal{U}^\dagger(t) \, \sigma_x \, \mathcal{U}(t) \\
&=  \sum_k J_k\left(\frac{\lambda}{\omega_s}\right) \left[\sigma_+\,  {\rm e}^{i\omega^{(k)} t}+ \sigma_-\,  {\rm e}^{-i\omega^{(k) t} t}\right]\, .
\end{split}
\end{equation}
However, for stronger WM-reservoir coupling (thermal contact) and/or long correlation time of the reservoir, this bare dynamics is influenced by higher order correlations between WM and reservoirs (see below). This includes higher order quanta exchange between the periodically driven TLS and the reservoir oscillator modes with frequencies around $\Omega_s$ and $\Omega_f$, respectively.

Figure \ref{fig3} displays the dynamics of the ground state population and the heat current $I_h(t)$ in the periodic steady state with period $\tau_s$. Although Redfield$^+$ predicts a somewhat higher population than HEOM, the heat currents obtained by the two methods  are in excellent agreement for the chosen parameters of relatively strong and fast driving (see Fig. \ref{fig4}).

Figure~\ref{fig5}a,b compares heat currents for a strictly Markovian treatment and the approximate non-Markovian approach Redfield$^+$ with exact data from HEOM. A resonance-like pattern is apparent when $\omega_s$ is tuned away from the regime of very slow (adiabatic) driving  to the fast driving regime. The location of these resonances is captured by the Markovian treatment which, however, completely fails to predict correct resonance heights in contrast to the accurate agreement between Redfield$^+$ and HEOM.
Remarkably, broad maxima in the heat currents occurring at higher driving frequencies are completely absent in the Markovian treatment, a clear evidence that reservoir feedback and non-Markovianity play a dominant role in this domain as we will discuss in detail below. This is also true when the reservoir temperatures are interchanged such that the high frequency bath becomes  the cold (hot) reservoir and the low-frequency reservoir the hot one. Then, as seen Fig.~\ref{fig6}, 
the agreement between Redfield$^+$ and HEOM is less good. 
We conclude that within the chosen parameter domain non-Markovian effects are prominent at all driving frequencies, but particularly at faster driving and lower temperatures for the fast bath. 

Even when the parameter set of Fig.~\ref{fig5} is adopted for the reservoirs, but the driving amplitude for the WM is substantially increased, see Fig.~\ref{fig7}, the performance of Redfield$^+$ remains acceptable with minor deviations from HEOM only in the resonance range at slower driving frequencies again with the overall tendency to yield smaller absolute values for heat currents.  

Based on the above analysis (and more systematic results that are not shown here), we conclude that for the setting studied here the Redfield$^+$ is quantitatively correct in the range of driving frequencies $0\leq \omega_s\lesssim \omega_0$ and for driving amplitudes $\lambda\lesssim |\Omega_f+\Delta_f/2-\omega_0|, |\Omega_s-\Delta_f/2-\omega_0|$ (so that $\omega(t)$ does not exceed the full bandwidths of the reservoirs during one period) as long as $\Omega_f/T$ is on the order of 1 with $T$ being the temperature of the high frequency reservoir (i.e.\ the reservoir memory time is not too strong).

\section{Sideband resonances}
\label{sec:sidebands}

In order to understand the behavior of the heat currents when the driving frequency $\omega_s$ is varied, we start with the range of slow to moderate driving, where, as we will show, weak coupling master equation predict at least qualitatively resonant behavior. This is no longer true for faster driving as we will discuss in the next Section.

In the regime of weak coupling he interaction between the WM and the reservoirs is governed by the real-valued transition rates  
\begin{equation}\label{eq:rates}
\Gamma_{0/1}^{(k)} = \frac{\lambda^2}{4\omega_s^2}\, \left[S_h(\mp\omega^{(k)}) + S_c(\mp\omega^{(-k)})\right]\nonumber\\
\end{equation}
which directly determine the heat fluxes (cf.~ Appendix).
Consequently, the reservoirs are  only probed at the central ($\omega^{(0)}=\omega_0$) and sideband ($\omega^{(k)}, |k|\neq 0$) resonances, where only the sidebands contribute to the heat current and the contribution of each sideband is weighted by the Bessel function $J_k(\lambda/\omega_s)$, see Eq. (\ref{Eq:quasi}). The underlying approximation requires that $\omega_s\ll \omega_0, \Omega_\alpha$. 

\begin{figure}
\centering
\includegraphics[width=12cm]{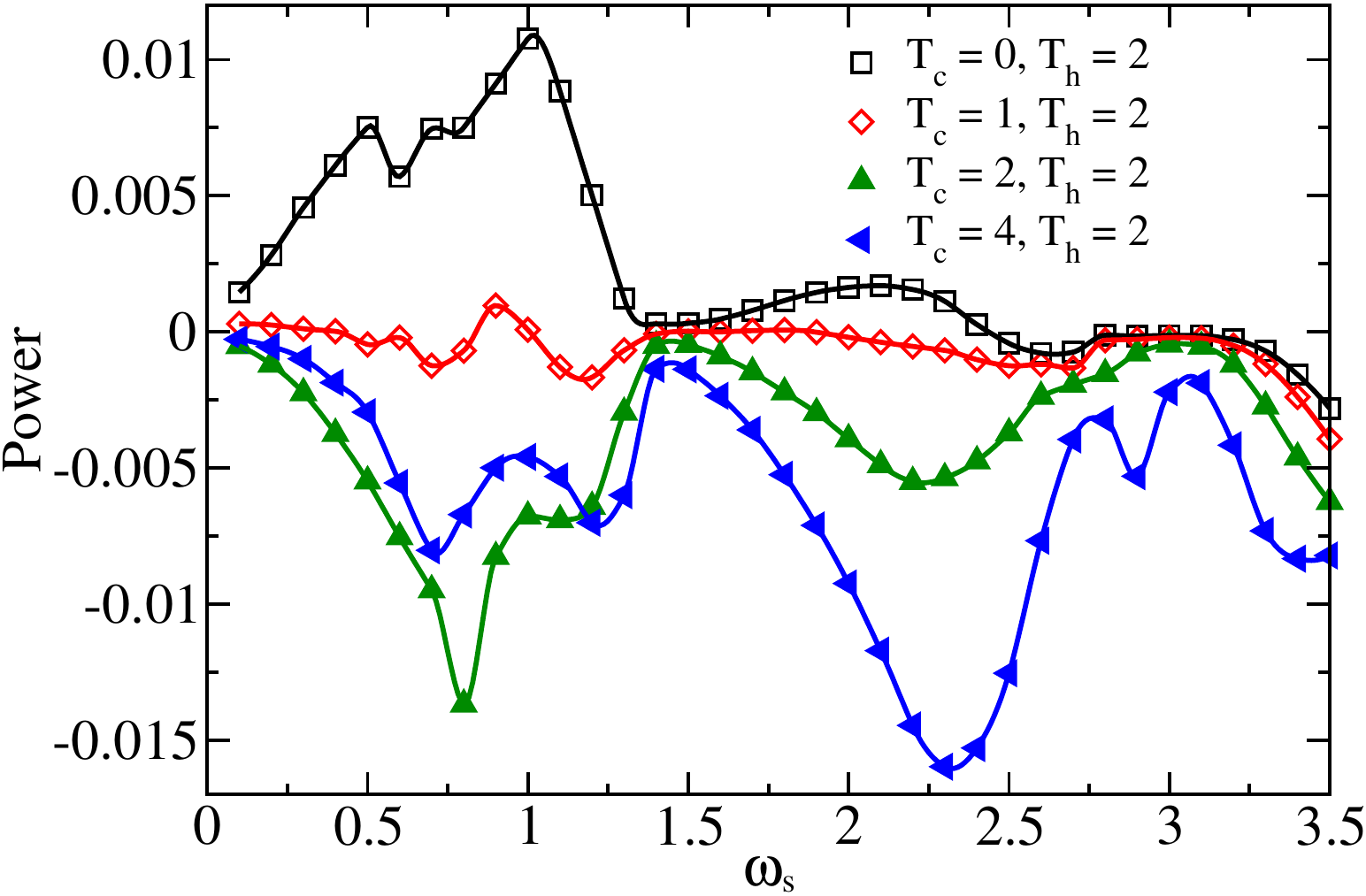}
\caption{Net power output as a function of the modulation frequency for various temperature gradients with $T_c$ allocated to the slow reservoir and $T_h$ to the fast one. The simulation parameters are (in a.u.): $\omega_0 = 3.0$, $\lambda = 1.0$, $\kappa_{\alpha} = 1.0$,
$\Omega_{s} = 2.0$, and $\Omega_{f} = 4.0$.}
\label{fig8}
\end{figure}
This treatment leads to the prediction that for the distributions of Eq. (\ref{eq:sp}) around $\Omega_\alpha$, we expect a resonance-like pattern for the heat currents around $\omega^{(k)}\approx \Omega_\alpha$ and thus around the driving frequencies
\begin{equation}
\label{eq:resonance}
    \omega_s^{(k)}=\frac{|\Omega_\alpha-\omega_0|}{k}\ , \ k=1, 2, 3, \dots\, . 
\end{equation}
In Fig.~\ref{fig5} this Markovian prediction for the heat current is shown together with the prediction  from Redfield$^+$ and HEOM. For the parameters chosen ($\lambda=|\Omega_\alpha-\omega_0|=1$), all treatments yield pronounced resonances around $\omega_s^{(1)}=1$ (single-quantum exchange) and $\omega_s^{(1)}=1/2$ (two-quanta exchange). However, the accuracy of the Markovian description is rather poor: the precise location of the resonance is shifted from the Markovian resonance condition and the peak heights differ substantially from the exact ones. The Markovian approximation completely fails in the limit $\omega_s\to 0$. In a more accurate description,
higher order quanta-exchange resonances are blurred by the steep decrease of the heat currents towards $\omega_s\to 0$. Apparently, the resonances are  broadened by the finite bandwidths of the reservoirs of order $\Delta_\alpha/2k$, see Figs.~\ref{fig5}--\ref{fig7}.

Memory effects of the reservoir response become even more prominent when the temperature gradient is reversed as in Fig.~\ref{fig6}, where the high frequency reservoir has $T_c=0$. This induces much stronger memory effects (non-Markovian behavior) on time scales of order $\tau_s$ so that (i) resonances occur slightly away from $\omega_s^{(k)}$ and (ii) higher order system-reservoir correlations cannot be neglected. They broaden the resonances and increase their magnitudes, thus, demonstrating that 'deep' quantum effects {\em enhance} the heat transfer. With increasing driving amplitude, see Fig.~\ref{fig7}, $\omega(t)$ covers the full bandwidths of the reservoirs, i.e.\ ${\rm max}_t \{\omega(t)\}\approx \Omega_f+\Delta_f/2$, ${\rm min}_t\{\omega(t)\}\approx \Omega_s-\Delta_s/2$, so that resonances overlap. 

With respect to the heat power, we observe in Figs.~\ref{fig5} and \ref{fig6} the expected behavior of peaks at the sideband resonances according to the respective temperature gradients. This is shown in more detail in Fig.~\ref{fig8}, where the heat power turns from being positive to being negative when the temperature of the low frequency reservoir increases. Note that in all cases, finite heat power also appears outside the range, where sideband resonances exist ($\omega_s>\omega_s^{(1)}$), i.e.\ outside the Markovian domain. We will discuss this latter range in the following.

\section{Power boost by non-Markovianity for fast driving}
\label{sec:nonMarkov1}

We also observe broad extrema in the heat currents for modulation frequencies around $\omega_s=2$. To explain their nature, we consider the limit, where the two reservoirs collapse to single oscillator modes with frequencies $\Omega_\alpha$. Accordingly, one has 
\begin{equation}
    H_{\rm harmonic}(t) = \omega(t)\sigma_+\sigma_- + \sum_\alpha \left\{ \frac{p_\alpha^2}{2} + 
    \frac{1}{2}\Omega_\alpha^2 q_\alpha^2 - c_\alpha q_\alpha\, \sigma_x \right\} \;\;
\end{equation}
with the corresponding Heisenberg equations of motion  
\begin{subequations}\label{eq:eomsigq}
\begin{equation}
\Ddot{q}_\alpha(t) + \Omega_\alpha^2 q_\alpha(t) = c_\alpha \sigma_x(t) \;\;;
\end{equation}
\begin{equation}
    \dot{\sigma}_x(t) + \omega(t)\sigma_y(t) = 0 \;\;;
\end{equation}
\begin{equation}
    \dot{\sigma}_y(t) -\omega(t)\sigma_x(t) - 2 [c_f q_f(t)+c_s q_s(t)]\sigma_z(t)= 0 \;\;;
\end{equation}
\begin{equation}
    \dot{\sigma}_z(t) + 2[c_f q_f(t)+c_s q_s(t)]\sigma_y(t) = 0\;\;.
\end{equation}
\end{subequations}
One can iterate the above equations to obtain for the reservoir oscillator modes
\begin{equation}\label{Eq:pr}
\begin{split}
    \Ddot{q}_\alpha(t) + \Omega_\alpha^2 q_\alpha(t) + 2c_\alpha^2 &\int_0^tdu\int_0^{u}ds\,  \omega(s)\sigma_z(s)\, q_\alpha(s) \\
    &=-c_\alpha \int_0^tdu\, \omega(u)\int_0^{u}ds\,  [\omega(s)\sigma_x(s)+2 c_{\bar{\alpha}}q_{\bar{\alpha}}(s) \sigma_z(s)]\;.
\end{split}
\end{equation}
with the index $\bar{\alpha}=f, s$ for $\alpha=s, f$. To order $c_\alpha^2$, this equation describes linearly and parametrically driven harmonic systems while higher order couplings induce nonlinearities. 
\begin{figure}
\centering
\includegraphics[width=12cm]{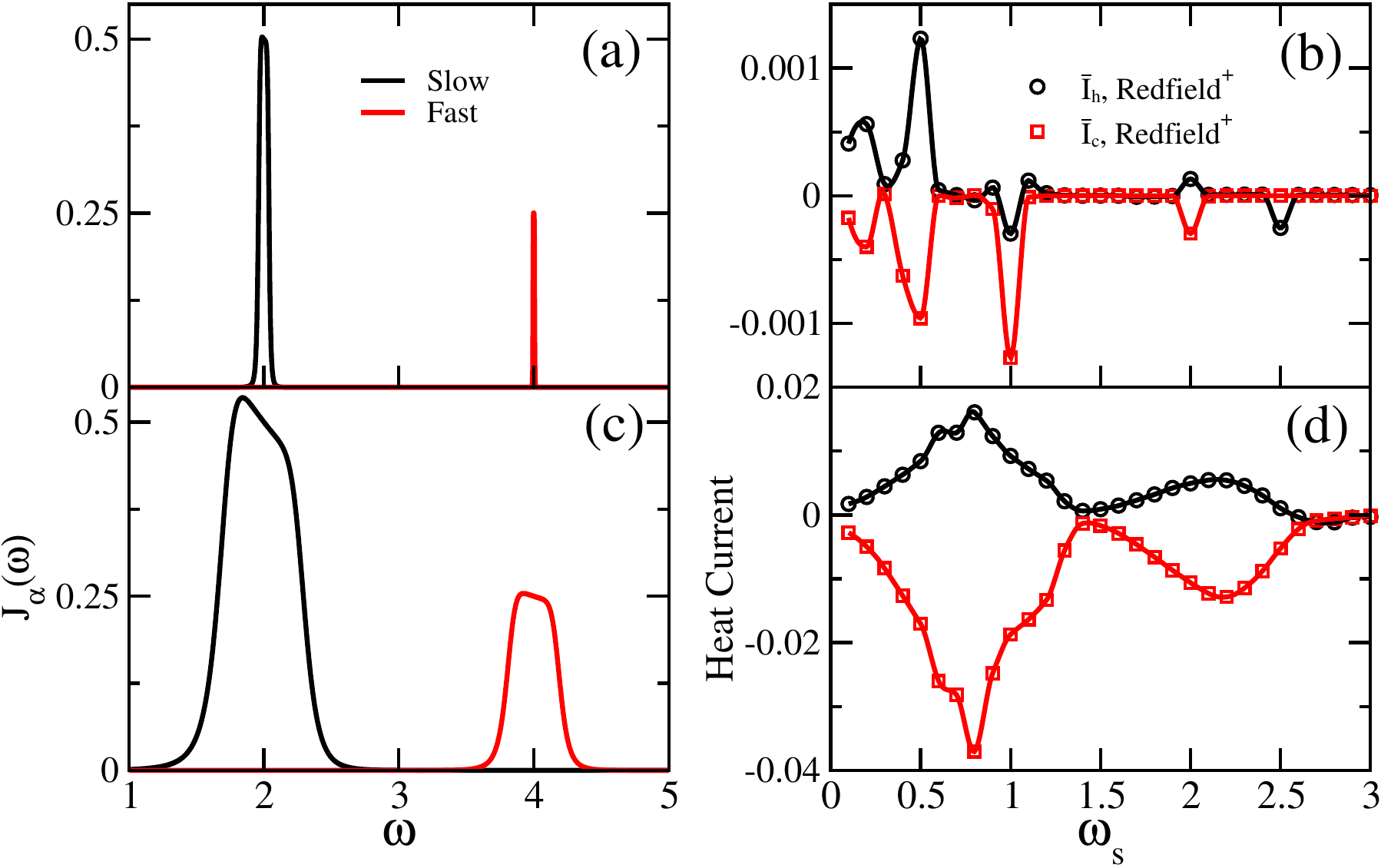}
\caption{Spectral distributions of slow and fast reservoirs  for narrow bandwidths (panel (a) given by Eq. (\ref{eq:sp-1})) and for broad bandwidth (panel (c) given by  Eq. (\ref{eq:sp})). Panel (b) and panel (d) show the corresponding heat current modulated by driving frequency $\omega_s$. The simulation parameters are (in a.u.): $\omega_0 = 3.0$, $\lambda = 1.0$, $\kappa_{\alpha} = 1.0$, ($\Omega_s, T_h) = (2.0,2.0)$, and ($\Omega_f,T_c) = (4.0,0.0)$.}
\label{fig9}
\end{figure}

The time dependent heat current follows according to (\ref{Eq:qhc}) from $I_\alpha(t)\propto c_\alpha \langle \sigma_y q_\alpha\rangle_t$. Since to leading order $\langle q_\alpha\rangle_\alpha=0$, the $c_\alpha$-dependent terms in Eq. (\ref{eq:eomsigq} c) for $\sigma_y(t)$  are relevant which implies $I_\alpha \propto c_\alpha^2 \langle q_\alpha(t)q_\alpha(s)\sigma_z(s)\rangle$.
To second order in the couplings,  $\sigma_z(t)$ in $I_\alpha$ carries the bare frequencies $\omega^{(k)}$ and one regains the resonance condition Eq.  (\ref{eq:resonance}) for the time averaged heat current. Beyond this approximation, the coupled dynamics in Eqs. (\ref{eq:eomsigq}) and (\ref{Eq:pr})  describes oscillatory behavior at frequencies $\omega^{(k)}, n \Omega_s, m \Omega_f$ ($n, m $ being integer) and their combinations, thus giving rise to beating between WM and the reservoirs as well as between the reservoirs. Hence, in the time averaged heat current we expect in the range $\omega_s>\omega_s^{(1)}$ leading order resonances behavior at the frequencies $\omega_s\approx \Omega_\alpha, 2\Omega_\alpha$, at $\omega_s=(\omega_0+ \Omega_\alpha)/2$ and $\omega_s\approx (\Omega_f\mp\Omega_s)/2$.
\begin{figure}
\centering
\includegraphics[width=12cm]{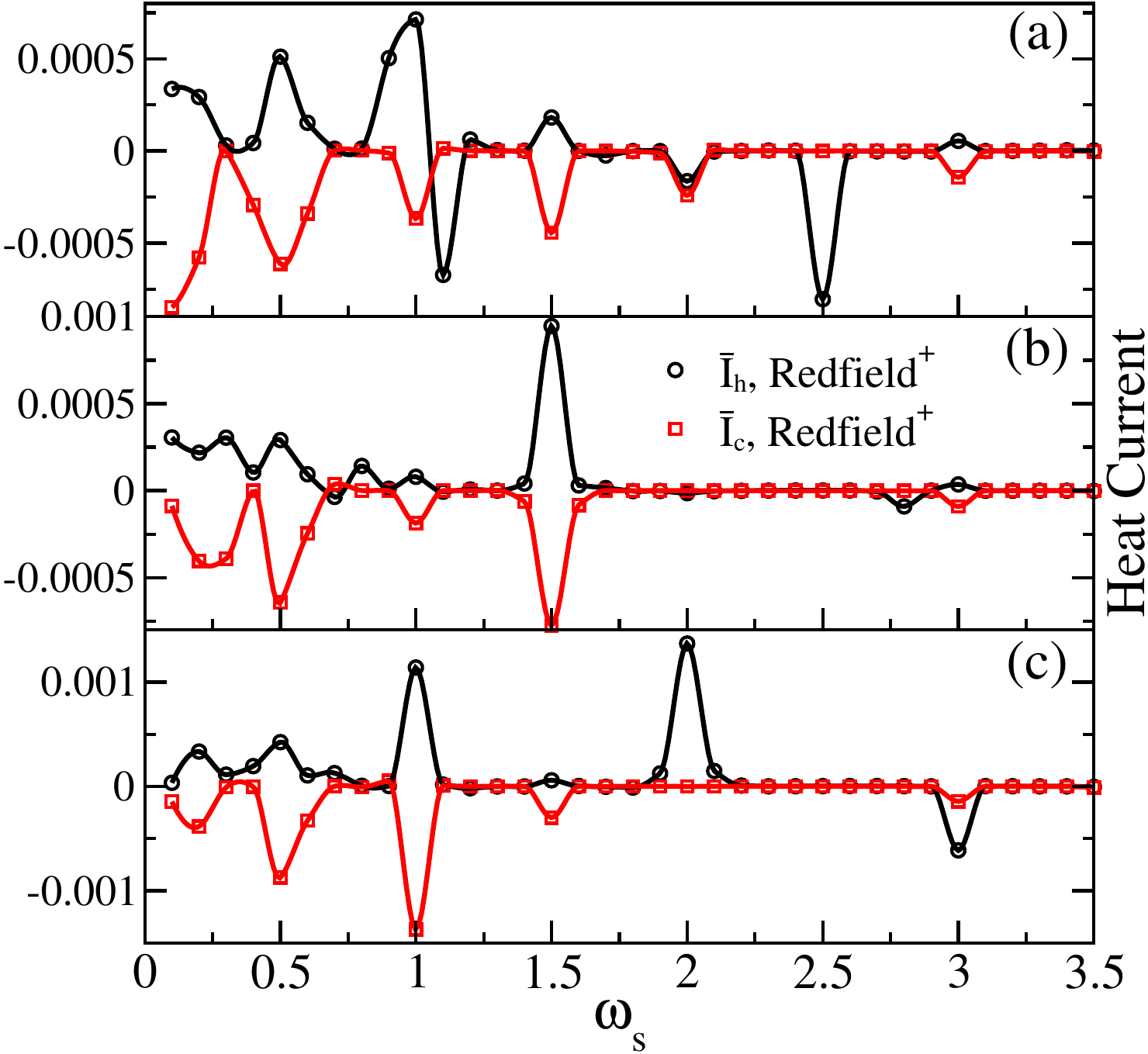}
\caption{Mean heat currents as a function of modulation frequency for narrow bandwidth reservoirs as in Fig.~\ref{fig9} but for (a) $\omega_0 = 3.0$, $\lambda = 2.0$; (b) $\omega_0 = 3.5$, $\lambda = 1.5$; (c) $\omega_0 = 4.0$, $\lambda = 2.0$. Other parameters are (in a.u.) $\kappa_{\alpha} = 1.0$, ($\Omega_{s},T_h) = (2.0, 2.0)$, and ($\Omega_{f},T_c) = (5.0, 0.0)$.}
\label{fig10}
\end{figure}

The above picture can be conveniently verified by considering narrow spectral distributions of the form
\begin{equation}\label{eq:sp-1}
  J_{\alpha}(\omega) = \frac{\kappa_{\alpha}\;\xi_{\alpha}^{20}\,\omega}
  {(\omega^4-\Omega_{\alpha}^4)^6 + \omega^2\xi_{\alpha}^{22}} \;\;, \ \alpha=s, f \;\;,
\end{equation}
which describe weakly damped oscillator modes with effective damping rates (spectral widths, cf. Fig. \ref{fig1}) $\Delta_\alpha\ll \Omega_\alpha$, see Fig.~\ref{fig9}(a). Corresponding results for the heat current are shown in Fig. \ref{fig9}(b): resonant structures are found for $\omega_s>1$ at $\omega_s=\Omega_s=2$ and $\omega_s=(\omega_0+\Omega_s)/2=2.5$. A broader spectral bandwidth of the individual reservoirs washes out individual resonances as seen in Figs. \ref{fig9}(c) and (d). The symmetry $|\omega_0-\Omega_f|=\omega_0-\Omega_s$ is broken in Fig. \ref{fig10} where extrema are in complete agreement with the above predictions.

\begin{figure}
\centering
\includegraphics[width=12cm]{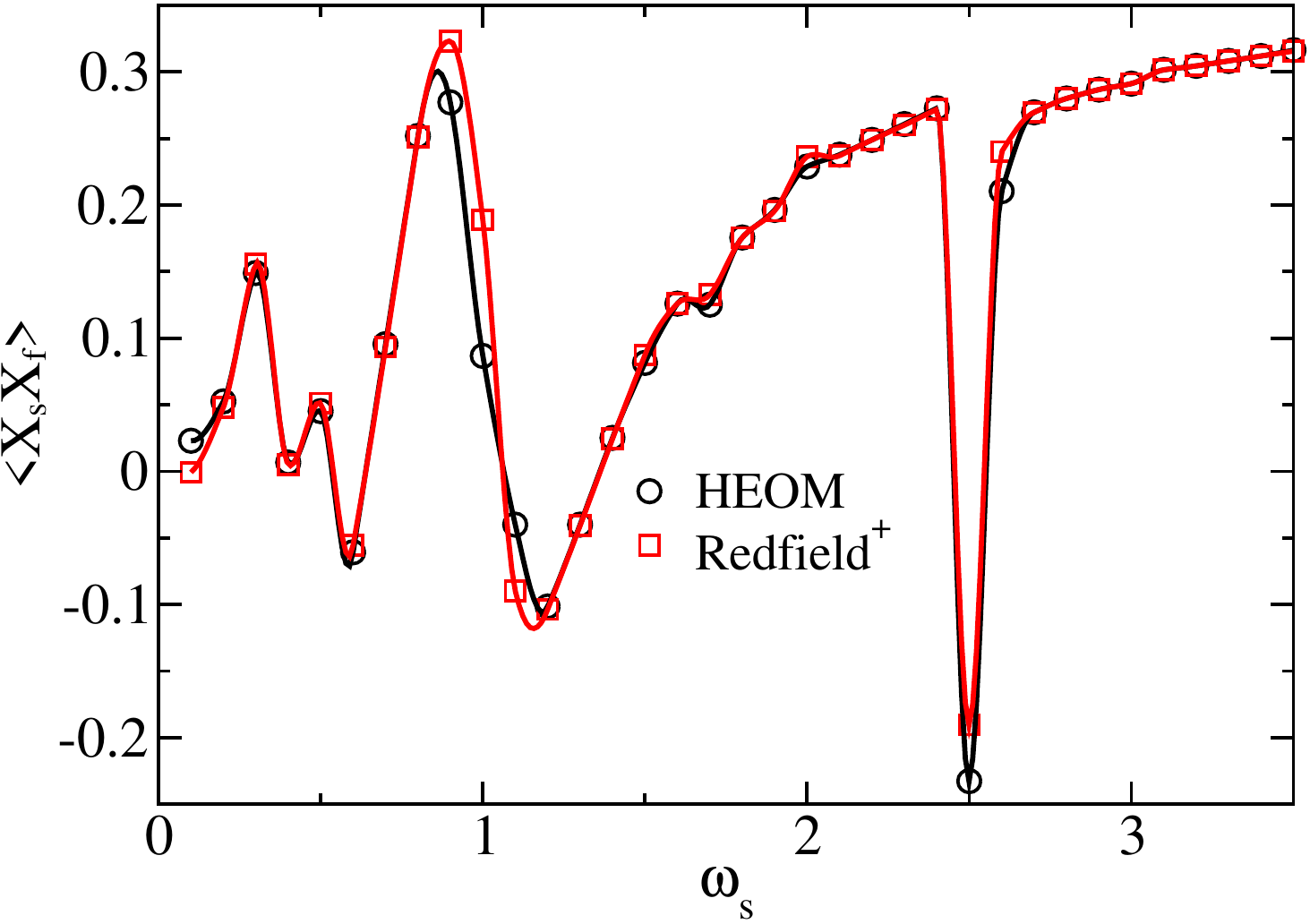}
\caption{Time averaged reservoir-reservoir correlation [cf. Eq. (\ref{Eq:xsxf})] as a function of the modulation frequency simulated by HEOM and  Redfield$^+$. The simulation parameters are the same as in Fig. \ref{fig10}(a).}
\label{fig11}
\end{figure}

Of particular interest are reservoir-reservoir correlations that we expect to emerge in the domain of strong non-equilibrium, i.e.\ for $\omega_s>\omega_s^{(1)}$. As Fig.~\ref{fig11} reveals, in the latter range these correlations are continuously built up with increasing modulation frequency approaching a constant level. In contrast, under slow and moderate modulation, these correlations only appear near resonances. Note that $\langle X_s X_f\rangle-$correlations do not exist in the Born-Markov approximation (which is a perturbative treatment up to order $c_\alpha^2$ in the secular approximation). These correlations reflect higher order system-bath contributions, at least of order $c_\alpha^4$, where they match the size of higher order contributions of auto-correlations (cf.~the normalization in Eq.  (\ref{Eq:xsxf})).

\begin{figure}
\centering
\includegraphics[width=12cm]{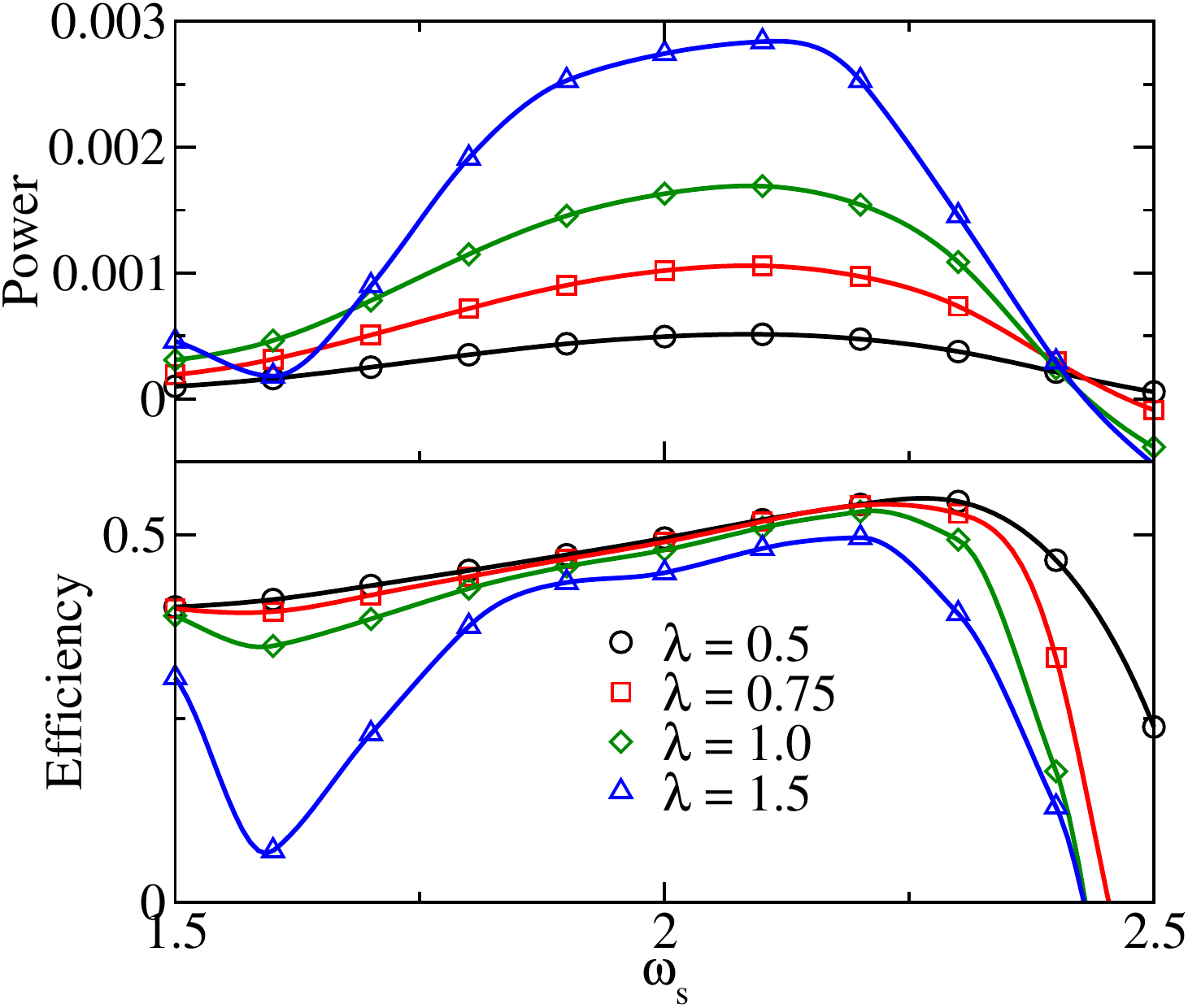}
\caption{Net power and efficiency versus the driving frequency for various driving amplitudes. The simulation parameters are (in a.u.): $\omega_0 = 3.0$, $\kappa_{\alpha} = 1.0$, ($\Omega_{s}$, $T_0$) = (2.0, 0.0), and 
($\Omega_{f}$, $T_h$) = (4.0, 2.0).}
\label{fig12}
\end{figure}

\begin{figure}
\centering
\includegraphics[width=12cm]{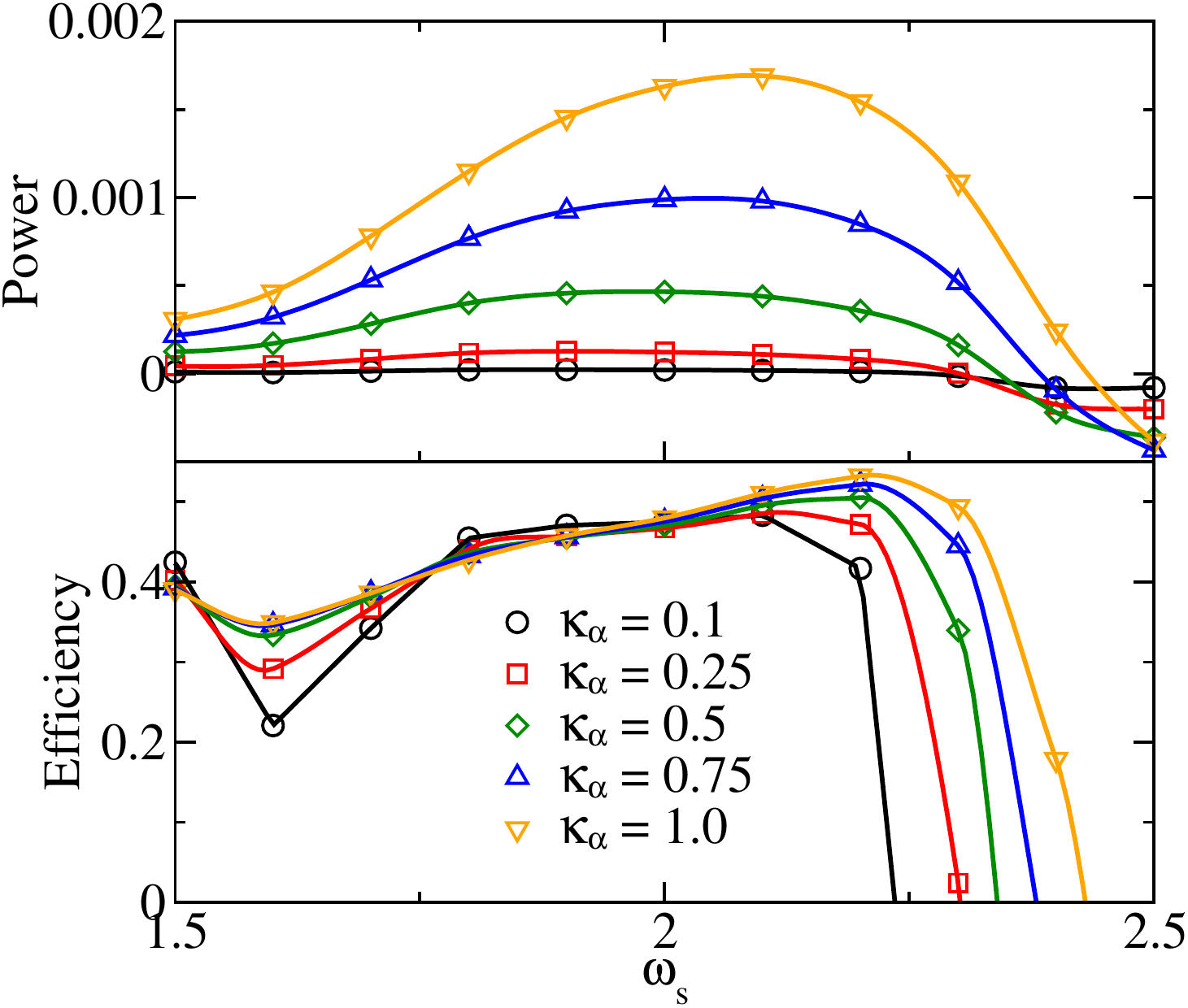}
\caption{Net power and efficiency versus the driving frequency  for various thermal coupling strengths of the WM with the reservoirs. The driving amplitude $\lambda = 1.0$, and other parameters are as in Fig.~\ref{fig12}.}
\label{fig13}
\end{figure}

Hence, from the above analysis we conclude that in addition to sideband resonances predicted from a Markovian treatment for slow to moderate driving, non-Markovian feedback effects dominate for faster modulation and lead to finite heat currents which would be absent otherwise. This effect consitutes non-Markovian power boost. This is further demonstrated in Figs.~\ref{fig12} and \ref{fig13}, which depict the dependence of power and efficiency on the modulation amplitude and the coupling to the reservoir. While the maximum power around $\omega_s\approx 2$ is sensitive to the driving strength, the efficiency is much less affected and even decreases for stronger modulation. A similar picture emerges for growing coupling: Stronger coupling provides more heat power but does not enhance the efficiency around the power maximum. Beyond the maximum of the power the engine turns into a heat dissipator with collapsing efficiency. Most remarkably, even without spectral overlap of the response frequency sidebands with the reservoir spectral density, at the modulation constant $\lambda=0.5$, for which the modulated $\omega(t)$ always remains within the bandgap between the reservoirs, we find substantial heat power with high efficiency. The reason is that response broadening due to the quantum time-energy uncertainty relation gives rise to the required spectral overlap in the non-Markovian anti-Zeno regime  \cite{mukherjee2020anti}. 

\section{Power boost through modulation of the thermal contact}
\label{sec:nonMarkov2}
So far we have considered a continuous coupling (denoted by scheme I) of the WM to its reservoirs. In order to further explore the machine performance in the non-Markovian domain, we extend the analysis to protocols which modulate periodically the WM-reservoir thermal contact (coupling strength). 
\begin{figure}
\centering
\includegraphics[width=12cm]{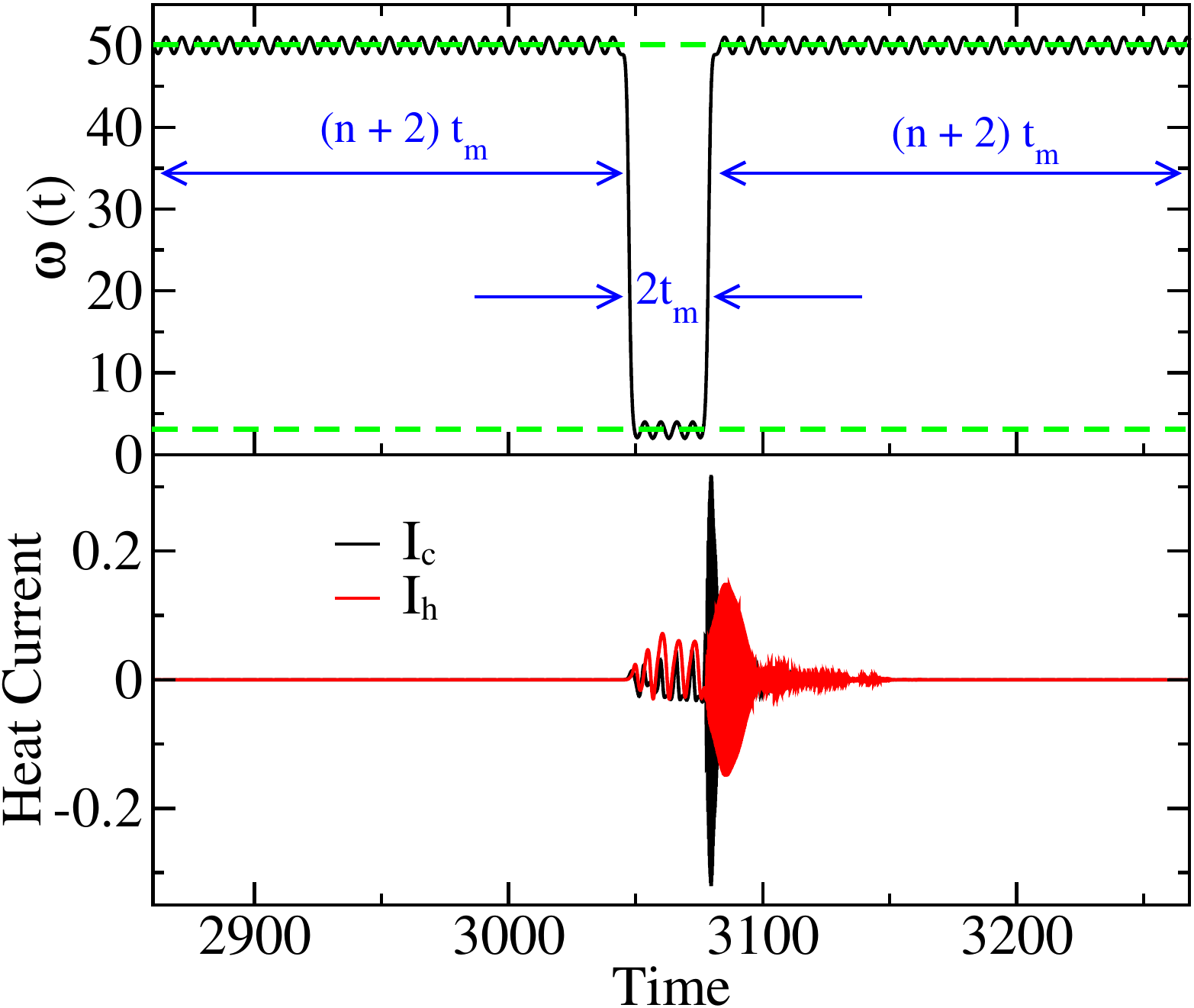}
\caption{Heat engine operated according to scheme III. Top: Modulation for one cycle of the heat engine  $\omega(t)=\bar{\omega}_0(t) +\lambda\cos(\omega_s t)$ with $\bar{\omega}_0$ as in Eq. \ref{Eq:wqt}. Time $t_m$ and integer $n$ are chosen properly such that $(n+2)t_m$ is on the order of the memory time of the reservoirs. Bottom: Dynamics of the heat current for $\omega_s=1, \lambda=1$; other parameter are as in Fig.~\ref{fig12}.}
\label{fig14}
\end{figure}
For this purpose, we choose two different protocols. Both start with an initialization step during which the engine approaches a steady state density from a factorized initial state. After this step thermal coupling is periodically switched on and off according to schemes: (II) Abrupt decoupling/coupling, as described in   \cite{mukherjee2020anti}, can be simulated by setting all ADOs in HEOM to either zero or non-zero. (III) Spectral decoupling/coupling by intermittently moving $\omega_0$ to a very high frequency above both reservoir bands and back to the reservoir gap, while still maintaning the modulation $\omega(t)$ in Eq. (\ref{Eq:wqt}) [see Fig.~\ref{fig14}]. Specifically, the decoupling cycle is described by a time dependence
\begin{equation}\label{Eq:w0t}
\begin{split}
    \bar{\omega}_0(t) =  \frac{\omega_1 + \omega_0}{2}+\frac{\omega_1-\omega_0}{2} &\left\{ \theta[(n+3)t_m/t_0 - t/t_0] \tanh \left[\frac{(n+2) t_m-t}{t_0} \right] \right. \\ 
    &\left.                     -          \theta[t/t_0 - (n+3)t_m/t_0] \tanh \left[\frac{(n+4) t_m-t}{t_0} \right]
                                            \right\} \;\;  
\end{split}                                            
\end{equation}
with $\theta(t)$ denoting the Heaviside step function and $\omega_1\gg \omega_0, \Omega_\alpha, \lambda$ being the off-resonance transition frequency. The time $t_0$ is chosen such that $t_m/t_0\gg 1$ so that, effectively, the WM is able to exchange heat with the reservoirs only during time spans of duration $2t_m$ while for times spans of duration $(n + 2)t_m$ it is decoupled. The cycle time is thus $\tau_C=(2n+6)t_m$ with an integer $n$ such that $(n+2)t_m$ is on the order of the memory time of the reservoirs $\tau_R$.

\begin{figure}
\centering
\includegraphics[width=12cm]{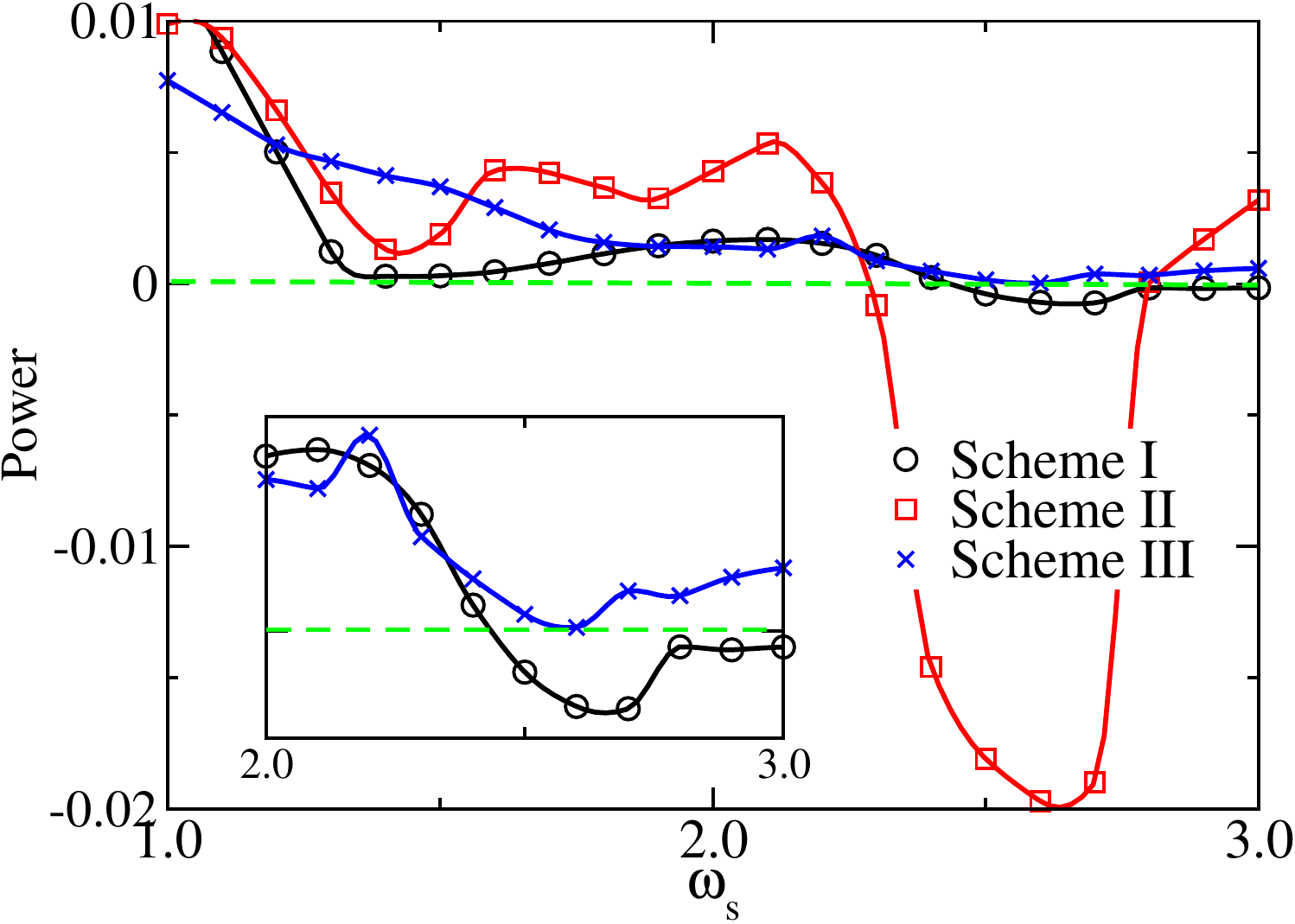}
\caption{Power boost due to non-Markovianity. Net power versus the driving frequency for the segmental driving protocols [scheme II (red)=abrupt decoupling, scheme III (blue)=continuous decoupling] compared to continuous driving (scheme I, black) for a cycle time $\tau_C=5 \tau_s$,  $\lambda=1$ and parameters $n = 10$, $t_0 = 1$, $\omega_1 = 50$, and $\omega_0 = 3$ for scheme III, see Eq. (\ref{Eq:w0t}); other parameters are as in Fig.~\ref{fig12}.
Since we infer the per-cycle work from both heat currents and the first law, additional work associated with (de)coupling~\cite{wiedmann2020non,wiedmann2021non,carrega2022engineering} is included.
A Markovian treatment predicts zero power (dashed line) and the inset shows the high frequency domain.}
\label{fig15}
\end{figure}

The resulting time dependent heat currents are depicted in Fig.~\ref{fig14}, lower panel. Note that the strongly oscillating part during and after the switch from $\omega_0\to \omega_1$ averages basically to zero so that the net heat is exchanged only when the transition frequency is modulated around $\omega_0$. The power performance is shown in Fig.~\ref{fig15} for both schemes II and III compared to a continuous process (scheme I, no active decoupling). We recall that a Markovian treatment predicts zero power in this case.

Both segmental schemes II and III lead to an enhanced power in the regime around $\omega_s=2$, where heat currents are now averaged over a full cycle. 
The abrupt decoupling scheme II turns the engine into a dissipator and then again into a heat engine for higher frequencies, while in the continuous decoupling/coupling scheme III it operates as heat engine throughout the range of modulation frequencies. While scheme II is more of an idealized thought experiment, scheme III can be implemented experimentally. 

These power boosts constitute a quantum advantage in agreement with the anti-Zeno effect \cite{mukherjee2020anti} that stems from the time-energy uncertainty relation under fast modulation of the system-reservoir interaction. Here the anti-Zeno effect arises in the strong-coupling regime and is evaluated non-perturbatively. 

\section{Conclusion}
\label{sec:conclusion}

This paper has studied a minimal model for a quantum heat engine in a non-trivial setting: By periodically modulating the transition frequency of a TLS acting as WM, this WM is in alternating spectral overlap with a cold or a hot reservoir with localized spectra that are separated (disjoint) by a bandgap. Such a continuously operating heat engine \cite{gelbwaser2013minimal,naseem2020minimal,gelbwaser2015thermodynamics,gelbwaser2014heat} has advantages over conventional stroke engines (for example the four-stroke Otto engine) since it does not require abrupt on-off switching of the coupling between WM and the reservoirs. As a result, the energetic and entropic cost of the switching is avoided. 

The present design extends the previous proposal \cite{mukherjee2020anti} to a much broader range of operational conditions, from weak to strong coupling of the WM and the reservoirs and weak to strong or slow to fast driving/ modulation of their coupling strength. Predictions have been obtained under any of the foregoing regimes without restrictions, based on the formally exact HEOM simulation technique and its comparison with perturbative treatments. While Markovian treatments predict vanishing heat flow at faster driving, the perturbative Redfield$^+$ can provide quantitatively correct predictions of the quantum heat engine performance for all driving frequencies. For slow and moderate driving the WM exchanges energy with thermal reservoirs via multi-sideband resonances.

The essential feature of the reservoir spectral functions considered here is the presence of a bandgap with spectrally abrupt edges,  because it gives rise   to strong-coupling effects \cite{kofman1994spontaneous}, and allows for disjoint hot and cold reservoir spectra, as required for efficient heat engines that operate continuously \cite{gelbwaser2013minimal,gelbwaser2015thermodynamics,ghosh2018two,mukherjee2020anti}.  As noted in the Introduction, such bandgap reservoirs can be realized in photonic crystals \cite{kofman1994spontaneous,lambropoulos2000fundamental} their phononic analogs \cite{pascal2011circuit,schwab2000measurement,chang2006solid,pruttivarasin2011trapped,bouton2021quantum,kosloff2014quantum,anders2013thermodynamics,Perarnau2018strong,lobejko2020thermodynamics,ghosh2017catalysis,naseem2020minimal}, and are envisaged also in superconducting circuits. Although the exact spectral shape of the reservoirs outside bandgaps is of no qualitative importance, we note that reservoir spectra and the resulting energy (Lamb) shifts can be engineered using the principles discussed in refs. \cite{naseem2020minimal,senior2020heat,meschke2006single}.

The highlight finding has been the quantum anti-Zeno advantage  of the thermal machine for both continuous and segmental modulation in the deep non-Markovian regime. In the former case, strong bath feedback emerges due to non-equilibrium processes, while in the latter case, strong memory effects govern the quantum dynamics. The latter modulation protocol can be implemented in actual experimental settings.

\section*{Acknowledgment}
We thank Michael Wiedmann, Ronzani Alberto, and Jukka P. Pekola for fruitful discussion. M. X. acknowledges support by the state of Baden-Württemberg through bwHPC (JUSTUS 2 cluster). This work has been supported by IQST and the German Science Foundation (DFG) under AN336/12-1 (For2724). G.K. acknowledges support of the PACE IN Quantera project, the ISF, and the NSF-BSF.

\section*{Author contributions}
M. X. performed numerical simulations. All authors have been involved in model setting, results analysis, discussion of scientific results and in the writing of the manuscript.

\section*{Data availability}
The data that support the figures within this article are available from the corresponding author upon reasonable request.

\section*{Appendix: Heat flux in Born-Markov approximation}

The population dynamics of a driven two level system interacting with a bandgap reservoir as considered in the main text is governed by \cite{mukherjee2020anti}
\begin{equation}
\begin{split}
    \frac{d}{dt} P_0(t) &= \Gamma_0[1-P_0(t)] - \Gamma_1 P_0(t)  \\
                      &= -[\Gamma_0 + \Gamma_1]P_0(t) + \Gamma_0 \;\;,
\end{split}
\end{equation}
and $P_1(t)=1-P_0(t)$.  Asymptotically for $\dot{P}_i(t)=0$, this equation can be solved as 
\begin{equation}
P_i=\frac{\Gamma_i}{\Gamma_0+\Gamma_1}\,\;, i=0, 1
\end{equation}
with transition rates $\Gamma_{0/1}=\sum_k\Gamma_{0/1}^{(k)}$, where 
\begin{equation}
    \Gamma_{0/1}^{(k)} = \frac{\lambda^2}{4\omega_s^2}\, \left[S_h(\mp\omega^{(k)}) + S_c(\mp\omega^{(-k)})\right]\, .
\end{equation}
Accordingly, the heat currents in steady state are given by
\begin{subequations}
\begin{equation}
    I_h(\omega_s) = \frac{\lambda^2}{4\omega_s^2}\, \sum_k \omega^{(k)}\, S_h(\omega^{(k)})\, \frac{e^{-\beta_h\omega^{(k)}} - w}{w+1} \;\;;
\end{equation}
\begin{equation}
    I_c(\omega_s) = \frac{\lambda^2}{4\omega_s^2}\, \sum_k \omega^{(k)}\, S_c(\omega^{(-k)})\, \frac{e^{-\beta_c\omega^{(-k)}} - w}{w+1} \;\;,
\end{equation}
\end{subequations}
with $\omega^{(k)}$ as in Eq. (\ref{Eq:quasi}) and the population ratio
\begin{equation}
    w(\omega_s) = \frac{P_1}{P_0} = \frac{\Gamma_0}{\Gamma_1} \;\;.
\end{equation}

\bibliography{quantum}

\providecommand{\newblock}{}
\begin{thebibliography}{10}
\expandafter\ifx\csname url\endcsname\relax
  \def\url#1{{\tt #1}}\fi
\expandafter\ifx\csname urlprefix\endcsname\relax\def\urlprefix{URL }\fi
\providecommand{\eprint}[2][]{\url{#2}}

\bibitem{schwabl2006statistical}
Schwabl F 2006 {\em Statistical mechanics\/} (Springer Science \& Business
  Media)

\bibitem{benenti2017fundamental}
Benenti G, Casati G, Saito K and Whitney R~S 2017 {\em Phys. Rep.\/} {\bf 694}
  1--124

\bibitem{binder2018thermodynamics}
Binder F, Correa L~A, Gogolin C, Anders J and Adesso G (eds) 2018 {\em
  Thermodynamics in the Quantum Regime: Fundamental Aspects and New
  Directions\/} ({\em Fundamental Theories in Physics\/} vol 195) (Springer,
  Berlin)

\bibitem{klatzow2019experimental}
Klatzow J, Becker J~N, Ledingham P~M, Weinzetl C, Kaczmarek K~T, Saunders D~J,
  Nunn J, Walmsley I~A, Uzdin R and Poem E 2019 {\em Phys.~Rev.~Lett.\/} {\bf
  122} 110601

\bibitem{kosloff2013quantum}
Kosloff R 2013 {\em Entropy\/} {\bf 15} 2100--2128

\bibitem{kosloff2017quantum}
Kosloff R and Rezek Y 2017 {\em Entropy\/} {\bf 19} 136

\bibitem{weiss12}
Weiss U 2012 {\em Quantum dissipative systems\/} 4th ed (New Jersey: World
  Scientific)

\bibitem{gelbwaser2015thermodynamics}
Gelbwaser-Klimovsky D, Niedenzu W and Kurizki G 2015 {\em Advances In Atomic,
  Molecular, and Optical Physics\/} {\bf 64} 329--407

\bibitem{breuer02}
Breuer H~P and Petruccione F 2002 {\em The Theory of Open Quantum Systems\/}
  (New York: Oxford University Press)

\bibitem{carrega2016energy}
Carrega M, Solinas P, Sassetti M and Weiss U 2016 {\em Phys.~Rev.~Lett.\/} {\bf
  116} 240403

\bibitem{guarnieri2016energy}
Guarnieri G, Nokkala J, Schmidt R, Maniscalco S and Vacchini B 2016 {\em
  Phys.~Rev.~A\/} {\bf 94} 062101

\bibitem{aurell2017work}
Aurell E 2017 {\em Entropy\/} {\bf 19} 595

\bibitem{pezzutto2019out}
Pezzutto M, Paternostro M and Omar Y 2019 {\em Quantum Science and
  Technology\/} {\bf 4} 025002

\bibitem{uzdin2016quantum}
Uzdin R, Levy A and Kosloff R 2016 {\em Entropy\/} {\bf 18} 124

\bibitem{abiuso2019non}
Abiuso P and Giovannetti V 2019 {\em Phys.~Rev.~A\/} {\bf 99} 052106

\bibitem{mukherjee2020anti}
Mukherjee V, Kofman A~G and Kurizki G 2020 {\em Commun. Phys.\/} {\bf 3} 1--12

\bibitem{kofman2000acceleration}
Kofman A and Kurizki G 2000 {\em Nature\/} {\bf 405} 546--550

\bibitem{kofman2004unified}
Kofman A and Kurizki G 2004 {\em Phys.~Rev.~Lett.\/} {\bf 93} 130406

\bibitem{erez2008thermodynamic}
Erez N, Gordon G, Nest M and Kurizki G 2008 {\em Nature\/} {\bf 452} 724--727

\bibitem{wiedmann2020non}
Wiedmann M, Stockburger J~T and Ankerhold J 2020 {\em New.~J.~Phys.\/} {\bf 22}
  033007

\bibitem{wiedmann2021non}
Wiedmann M, Stockburger J~T and Ankerhold J 2021 {\em Eur. Phys. J. Spec.
  Top.\/}  1--7

\bibitem{uzdin2015equivalence}
Uzdin R, Levy A and Kosloff R 2015 {\em Physical Review X\/} {\bf 5} 031044

\bibitem{ghosh2018two}
Ghosh A, Gelbwaser-Klimovsky D, Niedenzu W, Lvovsky A~I, Mazets I, Scully M~O
  and Kurizki G 2018 {\em Proc. Natl. Acad. Sci. USA\/} {\bf 115} 9941--9944

\bibitem{restrepo2016driven}
Restrepo S, Cerrillo J, Bastidas V~M, Angelakis D~G and Brandes T 2016 {\em
  Phys.~Rev.~Lett.\/} {\bf 117} 250401

\bibitem{tanimura89}
Tanimura Y and Kubo R 1989 {\em J. Phys. Soc. Jpn.\/} {\bf 58} 101

\bibitem{kato2016quantum}
Kato A and Tanimura Y 2016 {\em J.~Chem.~Phys.\/} {\bf 145} 224105

\bibitem{Meng2021}
Xu M, Stockburger J and Ankerhold J 2021 {\em Phys. Rev. B\/} {\bf 103} 104304

\bibitem{motz2018rectification}
Motz T, Wiedmann M, Stockburger J~T and Ankerhold J 2018 {\em New.~J.~Phys.\/}
  {\bf 20} 113020

\bibitem{stockburger02}
Stockburger J~T and Grabert H 2002 {\em Phys.~Rev.~Lett.\/} {\bf 88} 170407

\bibitem{yang2020heat}
Yang C~H and Wang H 2020 {\em Entropy\/} {\bf 22} 1099

\bibitem{velizhanin2008heat}
Velizhanin K~A, Wang H and Thoss M 2008 {\em Chem.~Phys.~Lett.\/} {\bf 460}
  325--330

\bibitem{wang03}
Wang H and Thoss M 2003 {\em J.~Chem.~Phys.\/} {\bf 119} 1289--1299

\bibitem{esposito2015quantum}
Esposito M, Ochoa M~A and Galperin M 2015 {\em Phys.~Rev.~Lett.\/} {\bf 114}
  080602

\bibitem{esposito2015nature}
Esposito M, Ochoa M~A and Galperin M 2015 {\em Phys.~Rev.~B\/} {\bf 92} 235440

\bibitem{carrega2022engineering}
Carrega M, Cangemi L~M, De~Filippis G, Cataudella V, Benenti G and Sassetti M
  2022 {\em PRX Quantum\/} {\bf 3}(1) 010323
  \urlprefix\url{https://link.aps.org/doi/10.1103/PRXQuantum.3.010323}

\bibitem{yamamoto2018heat}
Yamamoto T, Kato M, Kato T and Saito K 2018 {\em New.~J.~Phys.\/} {\bf 20}
  093014

\bibitem{gull11}
Gull E, Millis A~J, Lichtenstein A~I, Rubtsov A~N, Troyer M and Werner P 2011
  {\em Rev.~Mod.~Phys.\/} {\bf 83} 349

\bibitem{rossnagel2016single}
Ro{\ss}nagel J, Dawkins S~T, Tolazzi K~N, Abah O, Lutz E, Schmidt-Kaler F and
  Singer K 2016 {\em Science\/} {\bf 352} 325--329

\bibitem{cottet2017observing}
Cottet N, Jezouin S, Bretheau L, Campagne-Ibarcq P, Ficheux Q, Anders J,
  Auff{\`e}ves A, Azouit R, Rouchon P and Huard B 2017 {\em Proc. Natl. Acad.
  Sci. USA\/} {\bf 114} 7561--7564

\bibitem{pekola2015towards}
Pekola J~P 2015 {\em Nat. Phys.\/} {\bf 11} 118--123

\bibitem{ronzani2018tunable}
Ronzani A, Karimi B, Senior J, Chang Y~C, Peltonen J~T, Chen C and Pekola J~P
  2018 {\em Nat. Phys.\/} {\bf 14} 991--995

\bibitem{senior2020heat}
Senior J, Gubaydullin A, Karimi B, Peltonen J~T, Ankerhold J and Pekola J~P
  2020 {\em Commun. Phys.\/} {\bf 3} 1--5

\bibitem{meschke2006single}
Meschke M, Guichard W and Pekola J~P 2006 {\em Nature\/} {\bf 444} 187--190

\bibitem{pascal2011circuit}
Pascal L~M~A, Courtois H and Hekking F~W~J 2011 {\em Phys.~Rev.~B\/} {\bf 83}
  125113

\bibitem{schwab2000measurement}
Schwab K, Henriksen E, Worlock J and Roukes M~L 2000 {\em Nature\/} {\bf 404}
  974--977

\bibitem{chang2006solid}
Chang C~W, Okawa D, Majumdar A and Zettl A 2006 {\em Science\/} {\bf 314}
  1121--1124

\bibitem{pruttivarasin2011trapped}
Pruttivarasin T, Ramm M, Talukdar I, Kreuter A and H{\"a}ffner H 2011 {\em
  New.~J.~Phys.\/} {\bf 13} 075012

\bibitem{bouton2021quantum}
Bouton Q, Nettersheim J, Burgardt S, Adam D, Lutz E and Widera A 2021 {\em Nat.
  Commun.\/} {\bf 12} 1--7

\bibitem{kosloff2014quantum}
Kosloff R and Levy A 2014 {\em Annu. Rev. Phys. Chem.\/} {\bf 65} 365--393

\bibitem{anders2013thermodynamics}
Anders J and Giovannetti V 2013 {\em New.~J.~Phys.\/} {\bf 15} 033022

\bibitem{Perarnau2018strong}
Perarnau-Llobet M, Wilming H, Riera A, Gallego R and Eisert J 2018 {\em Phys.
  Rev. Lett.\/} {\bf 120}(12) 120602
  \urlprefix\url{https://link.aps.org/doi/10.1103/PhysRevLett.120.120602}

\bibitem{lobejko2020thermodynamics}
{\L}obejko M, Mazurek P and Horodecki M 2020 {\em Quantum\/} {\bf 4} 375

\bibitem{ghosh2017catalysis}
Ghosh A, Latune C, Davidovich L and Kurizki G 2017 {\em Proc. Natl. Acad. Sci.
  USA\/} {\bf 114} 12156--12161

\bibitem{gelbwaser2013minimal}
Gelbwaser-Klimovsky D, Alicki R and Kurizki G 2013 {\em Phys.~Rev.~E\/} {\bf
  87} 012140

\bibitem{naseem2020minimal}
Naseem M~T, Misra A, M{\"u}stecaplio{\u{g}}lu {\"O}~E and Kurizki G 2020 {\em
  Phys. Rev. Research\/} {\bf 2} 033285

\bibitem{gelbwaser2015work}
Gelbwaser-Klimovsky D and Kurizki G 2015 {\em Sci. Rep.\/} {\bf 5} 1--6

\bibitem{gelbwaser2014heat}
Gelbwaser-Klimovsky D and Kurizki G 2014 {\em Phys.~Rev.~E\/} {\bf 90} 022102

\bibitem{liu2021periodically}
Liu J, Jung K~A and Segal D 2021 {\em Phys.~Rev.~Lett.\/} {\bf 127} 200602

\bibitem{kofman1994spontaneous}
Kofman A, Kurizki G and Sherman B 1994 {\em J. Mod. Opt.\/} {\bf 41} 353--384

\bibitem{lambropoulos2000fundamental}
Lambropoulos P, Nikolopoulos G~M, Nielsen T~R and Bay S 2000 {\em Rep. Prog.
  Phys.\/} {\bf 63} 455

\bibitem{tanimura06}
Tanimura Y 2006 {\em J. Phys. Soc. Jpn.\/} {\bf 75} 082001--082039

\bibitem{tanimura2020numerically}
Tanimura Y 2020 {\em J.~Chem.~Phys.\/} {\bf 153} 020901

\bibitem{kato15}
Kato A and Tanimura Y 2015 {\em J.~Chem.~Phys.\/} {\bf 143} 064107

\bibitem{song17a}
Song L and Shi Q 2017 {\em Phys.~Rev.~B\/} {\bf 95} 064308

\bibitem{feynman63}
Feynman R~P and Vernon F~L 1963 {\em Ann. Phys.\/} {\bf 24} 118

\bibitem{magazzu2017asymptotic}
Magazz{\`u} L, Denisov S and H{\"a}nggi P 2017 {\em Phys.~Rev.~A\/} {\bf 96}
  042103

\bibitem{magazzu2018asymptotic}
Magazz{\`u} L, Denisov S and H{\"a}nggi P 2018 {\em Phys.~Rev.~E\/} {\bf 98}
  022111

\bibitem{grifoni1998driven}
Grifoni M and H{\"a}nggi P 1998 {\em Phys. Rev.\/} {\bf 304} 229--354

\bibitem{magazzu2018probing}
Magazz{\`u} L, Forn-D{\'\i}az P, Belyansky R, Orgiazzi J~L, Yurtalan M, Otto
  M~R, Lupascu A, Wilson C and Grifoni M 2018 {\em Nat. Commun.\/} {\bf 9} 1--8

\bibitem{traversa2013generalized}
Traversa F~L, Di~Ventra M and Bonani F 2013 {\em Phys.~Rev.~Lett.\/} {\bf 110}
  170602

\bibitem{kosloff2014}
Levy A and Kosloff R 2014 {\em Europhys. Lett.\/} {\bf 107} 20004

\bibitem{paauw2009tuning}
Paauw F, Fedorov A, Harmans C~M and Mooij J 2009 {\em Phys. Rev. Lett.\/} {\bf
  102} 090501

\bibitem{tanimura14}
Tanimura Y 2014 {\em J.~Chem.~Phys.\/} {\bf 141} 044114

\bibitem{song15b}
Song L and Shi Q 2015 {\em J.~Chem.~Phys.\/} {\bf 143} 194106

\bibitem{echave92}
Echave J and Clary D~C 1992 {\em Chem.~Phys.~Lett.\/} {\bf 190} 225

\bibitem{jin08}
Jin J~S, Zheng X and Yan Y~J 2008 {\em J. Chem. Phys.\/} {\bf 128}
  234703--234717

\bibitem{tang15}
Tang Z, Ouyang X, Gong Z, Wang H and Wu J 2015 {\em J.~Chem.~Phys.\/} {\bf 143}
  224112

\bibitem{meier99}
Meier C and Tannor D 1999 {\em J.~Chem.~Phys.\/} {\bf 111} 3365

\bibitem{ishizaki05}
Ishizaki A and Tanimura Y 2005 {\em J. Phys. Soc. Jpn.\/} {\bf 74} 3131--3134

\bibitem{xu17}
Xu M, Song L, Song K and Shi Q 2017 {\em J.~Chem.~Phys.\/} {\bf 146} 064102

\bibitem{zhang2020hierarchical}
Zhang H~D, Cui L, Gong H, Xu R~X, Zheng X and Yan Y 2020 {\em J.~Chem.~Phys.\/}
  {\bf 152} 064107

\bibitem{ikeda2020generalization}
Ikeda T and Scholes G~D 2020 {\em J.~Chem.~Phys.\/} {\bf 152} 204101

\bibitem{yan2020new}
Yan Y, Xing T and Shi Q 2020 {\em J.~Chem.~Phys.\/} {\bf 153} 204109

\bibitem{tanimura90}
Tanimura Y 1990 {\em Phys.~Rev.~A\/} {\bf 41} 6676

\bibitem{shi09b}
Shi Q, Chen L~P, Nan G~J, Xu R~X and Yan Y~J 2009 {\em J. Chem. Phys.\/} {\bf
  130} 084105--084108

\bibitem{cui2019highly}
Cui L, Zhang H~D, Zheng X, Xu R~X and Yan Y 2019 {\em J.~Chem.~Phys.\/} {\bf
  151} 024110

\bibitem{shi2018efficient}
Shi Q, Xu Y, Yan Y and Xu M 2018 {\em J.~Chem.~Phys.\/} {\bf 148} 174102

\bibitem{borrelli2019density}
Borrelli R 2019 {\em J.~Chem.~Phys.\/} {\bf 150} 234102

\bibitem{dunn2019removing}
Dunn I~S, Tempelaar R and Reichman D~R 2019 {\em J.~Chem.~Phys.\/} {\bf 150}
  184109

\bibitem{yan2021efficient}
Yan Y, Xu M, Li T and Shi Q 2021 {\em J.~Chem.~Phys.\/} {\bf 154} 194104

\bibitem{frishman96}
Frishman E and Shapiro M 1996 {\em Phys.~Rev.~A\/} {\bf 54}(4) 3310--3321

\bibitem{thanopulos08}
Thanopulos I, Brumer P and Shapiro M 2008 {\em J.~Chem.~Phys.\/} {\bf 129}
  194104

\bibitem{xu2018convergence}
Xu M, Yan Y, Liu Y and Shi Q 2018 {\em J.~Chem.~Phys.\/} {\bf 148} 164101

\bibitem{trushechkin2019higher}
Trushechkin A 2019 {\em Lobachevskii J. Math.\/} {\bf 40} 1606--1618

\bibitem{zhu12}
Zhu L, Liu H, Xie W and Shi Q 2012 {\em J.~Chem.~Phys.\/} {\bf 137} 194106

\bibitem{duan2020unusual}
Duan C, Hsieh C~Y, Liu J, Wu J and Cao J 2020 {\em J.~Phys.~Chem.~Lett.\/} {\bf
  11} 4080--4085

\end{thebibliography}
\end{document}